\documentclass[12pt,showpacs,amsmath,amssymb]{revtex4}
\usepackage{bm}
\usepackage{dcolumn}
\usepackage{graphicx}
\usepackage{epstopdf}
\usepackage{color}
\usepackage{epsf}
\usepackage{epsfig}

\begin{document}

\title{Double-Kasner Spacetime: Peculiar Velocities and Cosmic Jets}

\author{C. Chicone}

\affiliation{Department of Mathematics and Department of Physics and Astronomy, University of Missouri, Columbia,
Missouri 65211, USA }

\author{B. Mashhoon}

\affiliation{Department of Physics and Astronomy, University of Missouri,
Columbia, Missouri 65211, USA}

\author{K. Rosquist}

\affiliation{Department of Physics, Stockholm University, 106 91 Stockholm, Sweden\\and\\
International Center for Relativistic Astrophysics Network (ICRANet), Piazza della Repubblica 10, 65122 Pescara, Italy}

\begin{abstract}
In dynamic spacetimes in which asymmetric  gravitational collapse/expansion is taking place, the timelike geodesic equation appears to exhibit an interesting property: Relative to the collapsing configuration, free test particles undergo gravitational ``acceleration" and form a double-jet configuration parallel to the axis of collapse. We illustrate this aspect of peculiar motion in simple spatially homogeneous cosmological models such as the Kasner spacetime. To estimate the effect of spatial inhomogeneities on cosmic jets, timelike geodesics in the Ricci-flat double-Kasner spacetime are studied in detail. While spatial inhomogeneities can significantly modify the structure of cosmic jets, we find that under favorable conditions the double-jet pattern can initially persist over a finite period of time for sufficiently small inhomogeneities. 
\end{abstract}

\maketitle
\section{introduction}
Astrophysical jets are generally associated with cosmic systems that have undergone some form of gravitational collapse. In fact, the direction of the double-jet configuration of high-energy outflows is commonly presumed to coincide with the axis of rotation of the central gravitationally collapsed source. Similar double-jet patterns have been found in the recent analytic studies of the motion of free test particles relative to comoving observers in certain time-dependent solutions of Einstein's gravitational field equations~\cite{1,two}. In the spatially homogeneous Kasner spacetime, for instance, gravitational collapse occurs along one spatial axis, while there is expansion along the other two spatial axes. It can be shown that free test particles exhibit peculiar acceleration up and down parallel to the collapse axis leading to a double-jet configuration. Indeed, in time, peculiar velocities diminish to zero along expanding directions and approach the velocity of light along the contracting direction~\cite{two}. It is therefore worthwhile to investigate the robustness of these results in the presence of spatial inhomogeneities as in the double-Kasner spacetime. We find that for small inhomogeneities the double-jet feature involving peculiar acceleration indeed persists over a \emph{finite} time interval. Furthermore, in view of the current interest in ``dark flow", this work contributes to the theoretical study of bulk flows associated with peculiar motions in spatially inhomogeneous and anisotropic cosmological models. 

Consider a spacetime in which the proper distance along one spatial axis decreases to zero asymptotically (i.e., for $t\to\infty$), while the proper distance along each of the other two spatial axes tends asymptotically to infinity. Let us introduce a family of \emph{fundamental} observers in such a spacetime with four-velocity vector field $U^\mu=(-g_{tt})^{-1/2}\delta ^\mu{}_{0}$ and orthonormal tetrad field $\lambda^\mu{}_{(\alpha)}$ with $\lambda ^\mu{}_{(0)}=U^\mu$. We consider only positive square roots throughout. It is natural to express the four-velocity vector of a free test particle, $u^\mu=dx^\mu/d\tau$, with respect to the fundamental observers via $u^\mu=u^{(\alpha)} \lambda^\mu{}_{(\alpha)}$. Here $\tau$ is the proper time along the test particle's world line. Studying $u^{(\alpha)}$ as a function of time, we have found that there are timelike geodesics moving up and down parallel to the axis of collapse such that $u^{(0)}$, the Lorentz factor of the free test particle relative to the observer family, tends to infinity as $t\to\infty$. The remarkable gravitational ``acceleration" to the speed of light is an observer-independent feature of the double-jet configuration. This feature was first demonstrated in a Ricci-flat rotating cylindrically symmetric spacetime~\cite{1}.

The timelike geodesic equation in the spacetime of Ref.~\cite{1} has the structure of gravitomagnetic jets; that is, there are special timelike geodesics for a fixed value of the cylindrical radial coordinate that propagate along helical paths up and down parallel to the axis of cylinder, which is also the axis of rotation for this gravitational field. The speeds of free test particles that follow the special geodesics with respect to the fundamental observers (that are spatially at rest) tend to the speed of light as $t\to\infty$. Gravitomagnetic jets are attractors.  

We emphasize that in Ref.~\cite{1}, the physical spacetime domain in which gravitomagnetic jets occur is free of matter and singularities. The boundaries of this domain are excluded due to the inadmissibility of circular cylindrical coordinates. It is most likely that without cylindrical symmetry, an analytic treatment of
gravitomagnetic jets would not have been possible. Indeed, the main result---namely, the existence of a gravitational mechanism for the acceleration of free test particles to the speed of light---has nothing to do with the physical limitations of cylindrical symmetry. This follows from our recent study of cosmic jets in Kasner spacetimes~\cite{two}; that is, we show that similar results hold for the standard Kasner models of anisotropic cosmology. 

The purely gravitational mechanism for jet formation that we describe in Ref.~\cite{two} works as gravitational collapse is taking place; once a relatively stationary situation is established after the collapse, our mechanism ceases to function. We expect that other, basically electromagnetic, 
mechanisms would then take over to \emph{maintain} the jet that has just been formed.

More precisely, the various rays of high-energy particles that our gravitational mechanism originally produces may get confined and collimated where magnetic fields are strong. The initial double-jet configuration is then sustained over time by the various MHD mechanisms that have been
discussed in connection with astrophysical jets. The end result very much depends on the MHD aspects of the astrophysical environment. Either our initial streams of high-energy particles simply disperse via collisions every which way as cosmic rays or are confined, collimated and sustained by the MHD environment, or a combination of both. The central ``engine", in this approach, is whatever mechanism that \emph{sustains} the jet in a given environment.

In previous work on tidal dynamics~\cite{2,3,4,5,6,7,8}, the relative behavior of a congruence of geodesics has been studied. Moreover, \emph{tidal gravitational acceleration} of ultrarelativistic particles has been considered in a general context but via \emph{geodesic deviation} within the small-deviation approximation~\cite{3,4,5,6,7,8}. To study tidal acceleration/deceleration, imagine a congruence of timelike geodesics in a gravitational field; then, choosing one path in the congruence as representing the world line of our reference observer, we can express the motion of the other members of the congruence with respect to the local frame of the fiducial observer. A naturally invariant description of relative motion can be obtained by establishing a Fermi coordinate system in the neighborhood of the reference observer. The equation of motion of a nearby geodesic with arbitrary relative velocity (less than $c$) with respect to the fiducial observer is the generalized Jacobi equation. Indeed, the generalized Jacobi equation in a Fermi system has been employed to illustrate the concept of ``critical speed" given by $c/\sqrt{s}$, where $s = 2$  for linearized gravity in Fermi normal coordinates---a review of this approach is contained in Ref.~\cite{8}. This circumstance is compared in Ref.~\cite{8} with electrodynamics ($s = 1$), and, as expected, it is found that the corresponding critical speed is $c$. When the relative speed of the neighboring geodesics exceeds the critical speed $c/\sqrt{2} \approx{0.7 c}$, gravitational tidal forces can behave in a manner that is contrary to Newtonian expectations. This can lead, in certain circumstances, to the tidal acceleration of ultrarelativistic (i.e., $v > c/\sqrt{2}$) particles. The resulting gravitational acceleration mechanism has been extensively discussed in the \emph{stationary} exterior field of Kerr black holes in connection with the speed of astrophysical jets~\cite{3,4,5,6,7,8}. Instead of a single fiducial observer involved in the spatial neighborhood of free test particles, however, one can consider a whole family of observers associated with the ambient medium that perform pointwise measurements along the path of the free particles. That is, instead of tidal acceleration, we are interested here in \emph{peculiar acceleration}, namely, the  acceleration of free test particles relative to the comoving observers of the ambient medium. Hence, we adopt a completely different approach here;  in the present treatment, we deal with exact pointwise calculations within the context of certain specific \emph{dynamic} spacetimes. 
The critical speed turns out to be $c$ in this exact treatment,  which proves
to be \emph{formally} similar to electrodynamics. 

To see how this comes about, imagine an arbitrary timelike geodesic world line $\mathcal{W}$ with unit tangent vector $u^\mu=dx^\mu/d\tau$, where $\tau$ is the proper time along $\mathcal{W}$. We use units such that $c=1$ in what follows; moreover, the signature of the metric is $+2$. Let $\Lambda^\mu{}_{(\alpha)}$ be a smooth orthonormal tetrad frame field associated with an \emph{arbitrary} congruence of fiducial observers. In particular, along $\mathcal{W}$ we can write $u^\mu=u^{(\alpha)}\Lambda^\mu{}_{(\alpha)}$. Clearly,    $u^{(\alpha)}=u^{\mu} \Lambda_\mu{}^{(\alpha)}$ is coordinate independent, but it does depend on the background congruence of fiducial observers.  In particular, any ``acceleration" or ``deceleration" that could be deduced from the study of  $u^{(\alpha)}$  would depend on the background observers, since a geodesic particle has indeed a vanishing acceleration vector; that is,    $A^\mu=D u^\mu/d\tau=0$. The covariant derivative of $u^\mu$ along $\mathcal{W}$ vanishes; hence,
\begin{align} \label{eq:3}
\frac{
du^{(\alpha)}} {d\tau} \Lambda^\mu{}_{(\alpha)}+u^{(\alpha)}\frac{
D\Lambda^\mu{}_{(\alpha)}
}{d\tau}=0.
\end{align}
The rate of variation of the tetrad frame along $\mathcal{W}$ can be expressed as
\begin{align} \label{eq:4}
\frac{D\Lambda^\mu{}_{(\alpha)}}{d\tau}=\Phi_{(\alpha)}{}^{(\beta)}\Lambda^\mu{}_{(\beta)},
\end{align}
where $\Phi_{(\alpha)(\beta)}$ is antisymmetric due to the orthonormality of the frame field.
It follows from Eqs.~\eqref{eq:3} and~\eqref{eq:4} that 
\begin{align}\label{eq:5}
\frac{du^{(\alpha)}}{d\tau}=\Phi^{(\alpha)}{}_{(\beta)}u^{(\beta)}.
\end{align}
This result is \emph{formally}  analogous to the Lorentz force law. Just as electromagnetic acceleration of charged particles can be investigated using the Lorentz force law, the gravitational acceleration of free test particles relative to the congruence of reference observers can in principle  be studied using Eq.~\eqref{eq:5}. At any given event along the geodesic, the other observer families that one could imagine are related to the fiducial family by the elements of the Lorentz group. It is therefore essential in the study of this type of gravitational ``acceleration" relative to an observer family to focus on observer-independent features. In fact, the divergence of  $u^{(0)}$, the Lorentz factor of the free test particle relative to the congruence under consideration,   along the trajectory is the observer-independent feature that is the main focus of our work. That is, the measured speed of the test particle approaches the speed of light and this is an invariant property that is independent of any observer as a consequence of local Lorentz invariance. This divergence occurs in Ref.~\cite{1} and is the central feature of gravitomagnetic jets.

In conformity with the standard practice in relativistic cosmology,  in this work we consider explicitly the motion of free test particles relative to the background comoving observers. Moreover, in the simple examples that we study in this paper, instead of solving Eq.~\eqref{eq:5}, it is more convenient first to solve the geodesic equation for $u^{\mu}$ and then investigate the behavior of
\begin{equation}\label{eq:p}
u^{(\alpha)}=u^\mu \lambda _\mu{}^{(\alpha)}:=\gamma(1,\mathbf{v}),
\end{equation}
where $\textbf{v}$ is the \emph{peculiar velocity} of the free test particle relative to the family of \emph{fundamental} observers and $\gamma$ is the corresponding Lorentz factor.

\section{Peculiar Velocities}

To illustrate the behavior of peculiar velocities in spatially homogeneous spacetimes, consider a Bianchi type I model with a metric of the form
\begin{align}\label{16}
ds^2=-dt^2+X^2 dx^2+Y^2dy^2+Z^2 dz^2, 
\end{align}
where $X$, $Y$ and $Z$ are  functions of time $ t$. This spacetime is spatially homogeneous with three spacelike commuting  Killing vector fields      $\partial_x$ ,          $\partial_y$ and         $\partial_z$~\cite{12,14}.

The gravitational field equations  are
\begin{align}\label{17}
R_{\mu\nu}-\frac{1}{2} g_{\mu\nu} R=8\pi G\, T_{\mu\nu}, 
\end{align}
where  $T_{\mu\nu}$         is due to the presence of a perfect fluid with density $\rho$     and pressure    $P$,
\begin{align}\label{18}
T_{\mu\nu}=(\rho+P) U_\mu U_\nu+P g_{\mu\nu},
\end{align}
and the cosmological constant is assumed to be zero.
It follows that in comoving coordinates with  $U^\mu   = \delta^\mu{}_{0}$, we have
\begin{align}\label{19}
\frac{\dot{X}\dot{Y}}{XY}+\frac{\dot{Y}\dot{Z}}{YZ}+\frac{\dot{Z}\dot{X}}{ZX}&= 8\pi G\rho,\\ 
\label{20} \frac{\ddot{Y}}{Y}+\frac{\ddot{Z}}{Z}+\frac{\dot{Y}\dot{Z}}{YZ}&= -8\pi GP,\\
\label{21} \frac{\ddot{Z}}{Z}+\frac{\ddot{X}}{X}+\frac{\dot{Z}\dot{X}}{ZX}&= -8\pi GP,\\
\label{22} \frac{\ddot{X}}{X}+\frac{\ddot{Y}}{Y}+\frac{\dot{X}\dot{Y}}{XY}&= -8\pi GP.
\end{align}

Detailed discussions of the solutions of these equations for dust can be found, for instance, in section 5.4 of Ref.~\cite{HE} and section 12.15 of Ref.~\cite{PK}. The timelike geodesics, each with four-velocity  $u^\mu   = dx^\mu/d\tau$,      are such that the components of $u^{\mu}$ along Killing vector fields are constants; hence,
\begin{align}\label{23}
X^2\frac{dx}{d\tau}=\mathcal{C}_1,\quad Y^2\frac{dy}{d\tau}=\mathcal{C}_2,\quad Z^2\frac{dz}{d\tau}=\mathcal{C}_3,
\end{align}
where $\mathcal{C}_1$,  $\mathcal{C}_2$  and $\mathcal{C}_3$ are constants of the motion. Moreover, since $u^\mu$    is a unit vector, we find, 
\begin{align}\label{24}
\frac{dt}{d\tau}=\Big(1+\frac{\mathcal{C}^2_1}{X^2}+\frac{\mathcal{C}^2_2}{Y^2}+\frac{\mathcal{C}^2_3}{Z^2}\Big)^{1/2},
\end{align}
where $\tau$ is assumed to increase with $t$ along the world line.

 When $ \mathcal{C}_1=\mathcal{C}_2=\mathcal{C}_3= 0$, we have $u^\mu  =  U^\mu$, so that these timelike geodesics coincide with the fundamental comoving observers. The orthonormal tetrad frame associated with particles of the ambient perfect fluid (``fundamental observers") is then given by  $\lambda^\mu{}_{(0)}=U^\mu$                 and
\begin{align}\label{25}
\lambda^\mu{}_{(1)}=(0,\frac{1}{X},0,0),\quad 
\lambda^\mu{}_{(2)}=(0,0,\frac{1}{Y},0),\quad 
\lambda^\mu{}_{(3)}=(0,0,0,\frac{1}{Z}).
\end{align}
We are interested in the peculiar velocity of a free test particle with $(\mathcal{C}_1  , \mathcal{C}_2  , \mathcal{C} _3 ) \ne  0$ relative to the ambient medium. It then follows from Eq.~\eqref{eq:p} that  $\gamma=dt/d\tau$, which is  given by Eq.~\eqref{24},  and
\begin{align}\label{26}
\gamma v_x=\frac{\mathcal{C}_1}{X},\quad 
\gamma v_y=\frac{\mathcal{C}_2}{Y},\quad 
\gamma v_z=\frac{\mathcal{C}_3}{Z}.
\end{align}

Let $m > 0$ be the mass of a free test particle and $\mathcal{P}^{(\alpha)} = m u^{(\alpha)}$ be its peculiar four-momentum. It is a consequence of Eq.~\eqref{26} that  in general along the $x$ axis, $\mathcal{P}_{(x)} \propto X^{-1}$, and similarly along the $y$ and $z$ axes. As is well known, a similar relation holds in the Friedmann-Lemaitre-Robertson-Walker (``FLRW") models and expresses the law of decay of peculiar velocities as the universe expands (see Appendix A). Suppose, however, that along one axis---say, the $x$ axis---the universe contracts such that $X \to 0$; then, $\mathcal{P}_{(x)} \to \infty$. It follows that in this case the peculiar speed of the free test particle approaches the speed of light. This issue will be treated in this section in a couple of special cases of interest.

 Let us first consider the behavior of timelike geodesics in the standard Kasner metric~\cite{9}
\begin{equation}\label{eq:6}
ds^2=-dt^2+ \Big(\frac{t}{t_0}\Big)^{2p_1}dx^2+ \Big(\frac{t}{t_0}\Big)^{2 p_2} dy^2+ \Big(\frac{t}{t_0}\Big)^{2 p_3} dz^2,
\end{equation}
\begin{equation}\label{eq:7}
p_1+p_2+p_3=p_1^2+p_2^2+p_3^2=1.
\end{equation}
This empty universe model is a solution of Eqs.~\eqref{19}--\eqref{22} with $\rho = P = 0$. Here $t_0$ is the cosmic time at the present epoch. Henceforth, we will measure time in units of $t_0$; therefore, unless specified otherwise, we will formally set $t_0 = 1$ in what follows. We assume that    $p_1<p_2<p_3$;                    that is,
\begin{equation}\label{eq:8}
-\frac{1}{3}\le p_1\le 0,\qquad 0\le p_2\le \frac{2}{3},\qquad \frac{2}{3}\le p_3\le 1.
\end{equation}
In this standard expanding anisotropic cosmological model [10, 15, 16], Cartesian coordinates are admissible for $t  \in (0,\infty)$,              $t = 0$  at the cosmological singularity                    and  $(-g)^{1/2}=t$.                 Integrating $d\tau = dt/\gamma$ along a timelike geodesic in the Kasner spacetime and choosing the integration constant such that $\tau = 0$ at $t = 0$, we find that in general as $t$ goes from 0 to $\infty$, $\tau$ monotonically increases from 0 to $\infty$ as well. Furthermore, as cosmic time $t$ increases, the Kasner universe contracts along the $x$ axis while expanding along the $y$ and $z$ axes. It follows that as  $t\to\infty$,  $v_x\to \mathcal{C}_1/|\mathcal{C}_1|$, $v_y\to 0$ and $v_z\to 0$.
However, for free test particles with $\mathcal{C}_1   = 0$, $ \mathbf{v}\to  0$   as $t \to \infty$ . 
Thus in general---that is, except for a set of measure zero---all timelike geodesics asymptotically form a double-jet configuration, relative to comoving observers, that is parallel to the axis of collapse. Moreover for $t\to0$, $v_x\to 0$ and $\gamma\to \infty$. Thus in general peculiar velocities diminish to zero along expanding directions and increase up to the speed of light along contracting directions. These results are all the more remarkable because---except for the sign of $\mathcal{C}_1$---they do not depend significantly on the initial conditions for geodesic motion. In particular, in expanding directions, the \emph{magnitudes} of the initial velocities of the free test particles do not affect the end result~\cite{two}.

    It is interesting to consider, in Kasner spacetime, null geodesics, each  with tangent vector $k^\mu=dx^\mu/d\zeta$, where  $\zeta$       is an affine parameter along the path. Then, as before, we have that
\begin{align}
\label{eq:13a}
t^{2p_1} \frac{dx}{d\zeta}=N_1,\quad t^{2p_2} \frac{dy}{d\zeta}=N_2,\quad t^{2p_3} \frac{dz}{d\zeta}=N_3,
\end{align}
where $N_1$, $N_2$ and $N_3$   are constants of the null geodesic motion.  From  $k^\mu   k_\mu   = 0 $,  we find
\begin{align}
\label{eq:13b}
\frac{dt}{d\zeta}=(N_1^2 t^{-2p_1}+N_2^2 t^{-2p_2}+N_3^2 t^{-2p_3})^{1/2},
\end{align}
where $t $ is assumed to increase with   $\zeta$   along the world line.  It follows from an explicit comparison of Eqs.~\eqref{eq:13a}--\eqref{eq:13b} with the corresponding relations for timelike geodesics that timelike geodesics approach null geodesics for $ t \to\infty$ as well as for $t \to 0$; indeed, these features have been discussed in detail in Ref.~\cite{two} in terms of the notions of speed-of-light attractor and repellor, respectively.

Let us next turn to an examination of timelike geodesics in the Einstein-de Sitter model, which is a solution of Eqs.~\eqref{19}--\eqref{22} for $P = 0$ and $\rho = 1/(6 \pi G t^2)$. The metric is
\begin{align}
\label{eq:16} ds^2 =-dt^2+a^2(t)(dx^2+dy^2+dz^2),
\end{align}
where $a(t)=(t/t_0)^{2/3}$ and $t_0$ denotes the present cosmic epoch. The universe in this model expands from a singular state at $t = 0$ to $t = \infty$. Similarly, along any timelike geodesic the proper time $\tau$ goes from $\tau = 0$ to $\tau = \infty$, once the integration constant is so chosen that $\tau = 0$ at $t = 0$. The peculiar momentum of a free particle, $\mathcal{P}(t)$, pointwise measured by a comoving observer that the particle passes at cosmic epoch $t$, is proportional to $1/a(t)$. Thus in time the peculiar velocities monotonically decay, so that the free particles tend to a state of rest relative to the Hubble flow for $t \to \infty$, while for $t \to 0$, $a \to 0$ and hence $\gamma \to \infty$.

\begin{figure}
\includegraphics[scale=1.15,angle=0]{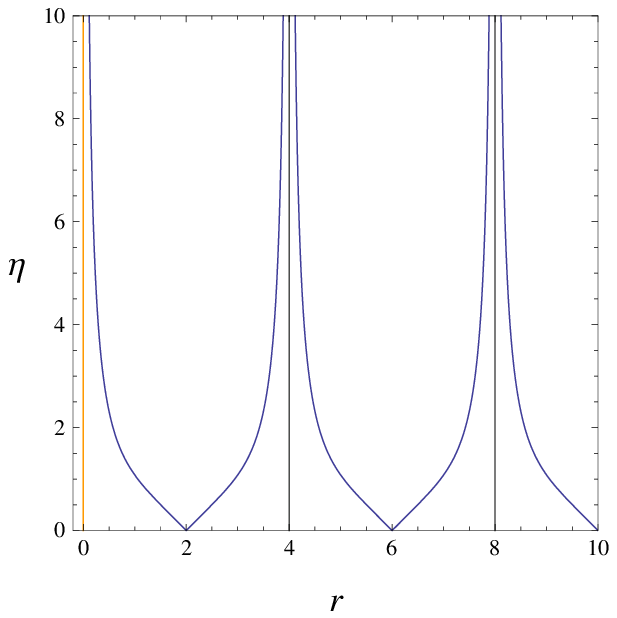}
\includegraphics[scale=1.15,angle=0]{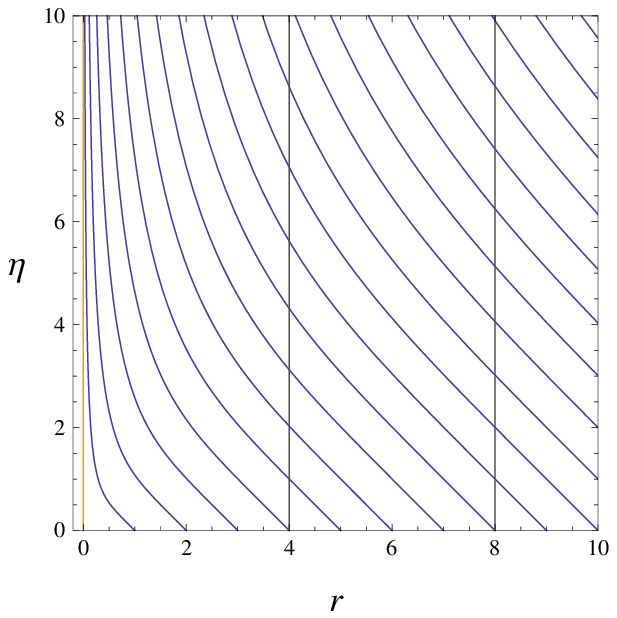}
\caption{Conformal time diagrams of the Einstein-de Sitter model showing free particle orbits. Lightlike geodesics would be at 45 degrees. Vertical lines are worldlines of fundamental observers. The curved lines are examples of non-comoving free orbits. The behavior as $t \rightarrow 0$ (corresponding to $\eta \rightarrow 0$) shows the (negative time) acceleration towards the speed of light relative to the fundamental observers (or equivalently the homogeneous hypersurfaces). Left panel: Orbits corresponding to one value of the parameter $b$ are shown. Right panel: The family of orbits which is attracted to the fiducial observer at $r=0$ in the limit $t \rightarrow 0$ is shown. These orbits have different values of $b$ corresponding to different initial distances from the fiducial observer.}  \label{fig:1}
\end{figure}

There is a particularly simple way of illustrating peculiar acceleration in the case of the Einstein-de Sitter universe. The idea is to display typical particle orbits using the standard conformal time diagram representation (see, e.g., Ref.~\cite{LL}). The conformal time $\eta$ is defined by the relation $dt=a(t)d\eta$. Then, $\eta=3 t_0^{2/3} t^{1/3}$, $\eta_0= 3t_0$ and the metric can be expressed in the form
\begin{equation}\label{K1}
   ds^2 = a^2(\eta)[-d\eta^2 + dr^2 + r^2(d\theta^2+\sin^2\!\theta d\phi^2)],
\end{equation}
where $a(\eta)= (\eta/\eta_0)^2$. We can focus on radial motion with respect to a fiducial fundamental observer without loss of generality because of isotropy and homogeneity. As follows from the above considerations, there is a constant of the motion given by
\begin{equation}\label{C_conf}
   \mathcal{C} = a^2 \frac{dr}{d\tau} \ , 
\end{equation}
where $\mathcal{C}^2 = \mathcal{C}_1^2 + \mathcal{C}_2^2 + \mathcal{C}_3^2$, so that the special value $\mathcal{C}=0$ corresponds to the fundamental comoving observers. Expressing Eq.~\eqref{24} for this case in conformal time and using Eq.~\eqref{C_conf} result in the equation for the orbits in the form
\begin{equation}\label{K2}
   \left( \frac{d\eta}{dr} \right)^2 = 1 + b^2 \eta^4,
\end{equation}
where $b= \mathcal{C}^{-1}\eta_0^{-2}$. This is an elliptic equation which can be reduced to Weierstrass form by using the auxiliary variables $\sigma= \eta^2$ and $u=br$. Then $\sigma(u)= \wp(u;h_2,h_3)$ is a Weierstrass elliptic function which solves the equation
\begin{equation}\label{K3}
   \sigma'(u)^2 = 4\sigma(u)^3 - h_2 \sigma(u) - h_3
\end{equation}
with $h_2= -4b^{-2}$ and $h_3=0$. Representative solutions for $\eta(r) = \sqrt{\sigma(br)}$ are shown in Fig.~\ref{fig:1}.

As can be understood from that figure, the fundamental observer congruence is stable in the future direction and it is attracting a space-filling family of orbits. It is also apparent that there is no acceleration in that direction. In the past direction, the situation is essentially the opposite, the flow is unstable and there is relative acceleration to the speed of light, $v \rightarrow 1$, as discussed above. In particular, in this case, the same family of orbits that is attracted in the future direction is repelled in the past direction.

We note in the examples above that the proper time can be chosen to be zero, say, at $t=0$, so that  the free particle encounters a curvature singularity when the peculiar acceleration to the speed of light occurs at a finite proper time along the geodesic. This circumstance appears necessary, since otherwise the geodesic could have been extended. The phenomenon of peculiar acceleration to the speed of light may possibly be used as a way of characterizing certain spacetime singularities. In short, if it occurs at a finite proper time of the relevant geodesic, it implies that the geodesic cannot be continued to that limit and beyond. This therefore indicates that the spacetime is timelike geodesically incomplete in the terminology of Ref.~\cite{HE}. It is clear that the singularity is nontrivial in the sense that it  cannot be removed by extending the spacetime in some way. The more detailed nature of such a  singularity is an interesting question that is beyond the scope of the present investigation.

We have thus far considered peculiar velocities in simple spatially homogeneous cosmological models. In the rest of this paper, we study peculiar velocities in a simple spatially inhomogeneous Kasner-like spacetime. It would be interesting to make a general study of peculiar velocities in spatially inhomogeneous cosmological models, as cosmological inhomogeneities tend to mimic dark energy---see~\cite{KB} and the references cited therein.

\section{double-Kasner spacetime}

To investigate the occurrence of cosmic double-jet configurations in the process of gravitational collapse, it is necessary to consider a realistic scenario involving a collapsing system. However, the actual physical situation involving a collapsing configuration of matter is extremely complicated to model properly; therefore, it is useful to consider exact solutions of Einstein's field equations that exhibit this property. Imagine, for simplicity, the Kasner metric~\eqref{eq:6} such that the metric coefficients also involve functions of spatial coordinates $(x,y,z)$.  While in this case, at every point in space, simultaneous collapse and expansion do occur along the spatial axes, they do so in a spatially inhomogeneous manner. This has the consequence that along a timelike geodesic in such a spacetime,  as $t$ varies monotonically with proper time $\tau$ by assumption, the peculiar motion involves metric functions that contain $x(\tau)$, $y(\tau)$ and $z(\tau)$ as well. The influence of spatial inhomogeneities may then totally dominate the collapse/expansion scenario. We show that this is indeed the case in certain situations via a detailed examination of the geodesics of  double-Kasner spacetime.

Consider, for simplicity,  the diagonal spacetime metric
\begin{equation}\label{N1}
ds^2=-e^{2q_0x}dt^2+ g_1^2 dx^2+g_2^2 dy^2+g_3^2 dz^2,
\end{equation}
where $q_0$ is a constant parameter and for $i = 1,2,3$, we define
\begin{equation}\label{N2}
g_i : = e^{q_i x}t^{p_i},
\end{equation}
so that spatial inhomogeneity is introduced here via exponential dependence of the metric coefficients upon the $x$ coordinate. 
Here $p_i$ are constants satisfying Eq.~\eqref{eq:7} as before and $q_i$ are given by
\begin{equation}\label{N3}
q_0 + q_2 + q_3 = q_1,\end{equation}  
\begin{equation}\label{N4}
q_0^2 + q_2^2 + q_3^2 = q_1^2,
\end{equation}  
\begin{equation}\label{N5}
q_0 (p_2+p_3) =  q_2 (p_2-p_1) + q_3 (p_3-p_1).
\end{equation} 
The coordinates $(t,x,y,z)$ in this \emph{double-Kasner spacetime} satisfy Lichnerowicz admissibility conditions once $t\in(0,\infty)$ and each spatial coordinate takes values in the open interval $(-\infty, \infty)$. There is a cosmological curvature singularity at $t = 0$ as in Kasner spacetime and  $(-g)^{1/2}=t \exp{(2q_1 x)}$; moreover, the spacetime contains two spacelike commuting Killing vector fields $\partial_y$ and $\partial_z$. The connection coefficients for metric~\eqref{N1} are given in Appendix B. 

It is possible to show that this spacetime is Ricci-flat. Following the work of Kasner~\cite{9}, it is natural to look for solutions of Einstein's field equations such that the metric tensor depends only on two coordinates, such as, for instance, $t$ and $z$. Assuming separability, such two-variable Kasner-type Ricci-flat solutions were found and  discussed in Ref.~\cite{HZ}.  In our case, the metric tensor depends on separable functions of $t$ and $x$ and is thus related to solutions of Ref.~\cite{HZ}; in fact, it belongs to Case B of Harris - Zund solutions~\cite{HZ}. Moreover, Harris and Zund showed that for normal parameters such solutions are algebraically general (Petrov type I), just as in the Kasner case~\cite{HZ}. The present Harris - Zund solution, essentially in the form of equations~\eqref{N1}--\eqref{N5}, was later given explicitly in Appendix 1 of Ref.~\cite{VD}. Further discussion of such multi-variable solutions is contained in section 17.3 of Ref.~\cite{12}.

It is a simple consequence of the algebraic relations~\eqref{N3}--\eqref{N5} that if $q_0 = 0$, then the spacetime is either flat or Kasner. Moreover, if $q_1 = 0$, then metric~\eqref{N1} simply reduces to the standard Kasner metric. Thus we assume in what follows that $q_0\ne0$ and $q_1\ne0$. Now let $\hat{q_0} = q_0/q_1$, $\hat{q_2} = q_2/q_1$ and $\hat{q_3} = q_3/q_1$; then, it follows from Eqs.~\eqref{N3} and~\eqref{N4} that the constants $\hat{\mathbf{q}}:=(\hat{q_0}, \hat{q_2}, \hat{q_3})$ satisfy the same algebraic relations given in Eq.~\eqref{eq:7} for $\mathbf{p}:=(p_1, p_2, p_3)$ and can be parameterized in the same way. The relationship between $\mathbf{p}$ and $\hat{\mathbf{q}}$, as discussed in detail in Appendix C, is marked by \emph{reciprocity}. Once $\mathbf{p}$ is chosen, then Eqs.~\eqref{N3}--\eqref{N5} determine $\hat{\mathbf{q}}$, thus leaving the metric still dependent upon one free parameter $q_1$, which could be positive or negative. Now under the parity transformation $x\to -x$, the metric remains invariant if the $q$'s are replaced by~$-q$'s, which leaves $\hat{\mathbf{q}}$ invariant; to remove the ambiguity in the sign of $q_1$ with no loss in generality, we assume henceforth that $q_1 > 0$. Moreover, we note that metric~\eqref{N1} is invariant under the exchange of $y$ with $z$ if simultaneously we exchange $p_2$ with $p_3$ and $q_2$ with $q_3$. This exchange invariance will be used in what follows. 

It is useful to introduce a dimensionless radial coordinate $\xi$ defined by
\begin{equation}\label{M1}
e^{q_1x} = \xi,
\end{equation} 
so that as $x \in (-\infty, \infty)$, we have $\xi \in (0, \infty)$. Under such a transformation, metric~\eqref{N1} takes the form
\begin{equation}\label{M2}
ds^2=-\xi^{2\hat{q_0}}dt^2+\frac{1}{q_1^2}t^{2p_1}d\xi^2+\xi^{2\hat{q_2}}t^{2 p_2} dy^2+\xi^{2\hat{q_3}}t^{2 p_3} dz^2.
\end{equation}
We recall that Kasner metric~\eqref{eq:6} is the standard timelike form of the Kasner solution, while its spacelike form may be written as~\cite{LL} 
\begin{equation}\label{M3}
ds^2=-x^{2p_1}dt^2+  dx^2+x^{2p_2} dy^2+x^{2p_3}dz^2.
\end{equation}
Here, Cartesian coordinates are admissible for $x  \in (0,\infty)$,              $x = 0$ is a curvature singularity and  $(-g)^{1/2}=x$.        \emph{It follows that double-Kasner metric~\eqref{M2} is indeed a nonlinear superposition of the timelike and spacelike forms of the Kasner metric.} In the form~\eqref{M2}, the double-Kasner metric belongs to Case A of the Harris - Zund classification~\cite{HZ}. We note that certain other ``double-Kasner" metrics have also been considered in Ref.~\cite{SK} and Ref.~\cite{RO}; in the former, in spacetimes of lower symmetry and in the latter, in 5D spacetime.

In the Kasner case, the three parameters cannot all be equal. If two are equal, then either $\mathbf{p} = (1,0,0)$, in which case the spacetime is flat or $\mathbf{p}= (-1/3, 2/3, 2/3)$, which corresponds to the non-flat plane-symmetric case. Otherwise, the three parameters are all different and take values within the open interval $(-1/3, 1)$, with one negative and two positive. Let us now consider in a similar way metric~\eqref{M2} by treating the special cases of $\mathbf{p}$ \emph{and} $\hat{\mathbf{q}}$. We first notice that $\mathbf{p}$ and $\hat{\mathbf{q}}$ cannot be equal in \emph{curved} spacetime. That is, $\mathbf{p}=\hat{\mathbf{q}}$ leads to $p_1=1$ or $\hat{q_0}=1$  or indeed both, since with $\mathbf{p} = (1,0,0)$, we find $\hat{\mathbf{q}} = (1,0,0)$; then, in terms of the radial coordinate $\xi$ defined in Eq.~\eqref{M1}, the double-Kasner metric reduces in this case to 
\begin{equation}\label{N6}
ds^2=- \xi^2 dt^2+\frac{1}{q_1^2} t^2 d\xi^2+ dy^2 + dz^2,
\end{equation}
which turns out to represent flat spacetime. Let us briefly digress here and mention a generalization of this case such that $p_1 = \hat{q_0}$; indeed, it can be shown that for a non-flat spacetime, we must in general assume that $p_1 \ne \hat{q_0}$. For instance, a double-Kasner spacetime with $\mathbf{p} = (0,1,0)$ and $\hat{\mathbf{q}} = (0,0,1)$ is flat. The other five special cases are \emph{not} flat and can be expressed as:
\begin{equation}\label{N7}
1. \quad \mathbf{p} = (0, 1, 0),  \quad  \hat{\mathbf{q}} = (\frac{2}{3}, \frac{2}{3}, -\frac{1}{3}),
\end{equation}
\begin{equation}\label{N8}
2. \quad \mathbf{p} = (-\frac{1}{3}, \frac{2}{3}, \frac{2}{3}),  \quad  \hat{\mathbf{q}} = (\frac{3}{7}, \frac{6}{7}, -\frac{2}{7}),
\end{equation} 
\begin{equation}\label{N9}
3. \quad \mathbf{p} = (\frac{2}{3}, -\frac{1}{3}, \frac{2}{3}),  \quad  \hat{\mathbf{q}} = (\frac{6}{7}, -\frac{2}{7}, \frac{3}{7}),
\end{equation} 
\begin{equation}\label{N10}
4. \quad \mathbf{p} = (\frac{3}{7}, \frac{6}{7}, -\frac{2}{7}),  \quad  \hat{\mathbf{q}} = (-\frac{1}{3}, \frac{2}{3}, \frac{2}{3}),
\end{equation} 
\begin{equation}\label{N11}
5. \quad \mathbf{p} = (\frac{6}{7}, -\frac{2}{7}, \frac{3}{7}),  \quad  \hat{\mathbf{q}} = (\frac{2}{3}, -\frac{1}{3}, \frac{2}{3}).
\end{equation}
Here we have taken into account the exchange invariance as well as the fact that $q_0\ne0$ and that for a non-flat spacetime $q_0\ne q_1$. 

Finally, let us note that there is a homothetic Killing vector field here just as in Kasner spacetime. We recall that in the non-flat case, $p_1 \ne 1$ and $q_0\ne q_1$. Consider the coordinate transformation $(t, x, y, z) \to (\tilde{t}, \tilde{x}, \tilde{y}, \tilde{z})$ given by
\begin{align}\label{N12}
t =e^{\omega_0} \tilde{t}, \quad x = \tilde{x} +\omega_1, \quad y = e^{\omega_2} \tilde{y}, \quad z = e^{\omega_3} \tilde{z},
\end{align}
where $\omega_0$ is a constant parameter and
\begin{equation}\label{N13}
\omega_1 = \frac{1-p_1}{q_1-q_0} \omega_0,
\end{equation}
\begin{equation}\label{N14}
\omega_2 = (q_0-q_2)\omega_1 +(1-p_2)\omega_0,
\end{equation}
\begin{equation}\label{N15}
\omega_3 = (q_0-q_3)\omega_1 +(1-p_3)\omega_0.
\end{equation}
Then metric~\eqref{N1} changes only by a constant factor, namely, 
\begin{equation}\label{N16}
ds^2= e^{2(\omega_0 + q_0 \omega_1)}d\tilde{s}^2.
\end{equation}
The homothetic generator can be easily deduced from the above considerations. 

\section{singularities of the double-Kasner spacetime}
A Ricci-flat spacetime has four algebraically independent scalar polynomial curvature invariants. They can be expressed as (cf. chapter 9 of~\cite{12})
\begin{align}
\label{S1} \mathcal{I}_1&=R_{\mu\nu\rho\sigma}R^{\mu\nu\rho\sigma}-iR_{\mu\nu\rho\sigma}R^{*\mu\nu\rho\sigma}, \\
\label{S2} \mathcal{I}_2&=R_{\mu\nu\rho\sigma}R^{\rho\sigma\alpha\beta}R_{\alpha\beta}{}^{\mu\nu}+iR_{\mu\nu\rho\sigma}R^{\rho\sigma\alpha\beta}{R^*}_{\alpha\beta}{}^{\mu\nu}.
\end{align}
For the double-Kasner spacetime represented by metric~\eqref{N1}, $\mathcal{I}_1$ and $\mathcal{I}_2$ are both real and are given by
\begin{equation}
\label{S3} \mathcal{I}_1 = - 16 p_1p_2p_3t^{-4}\xi^{-4\hat{q_0}} -16q_0q_1q_2q_3t^{-4p_1}\xi^{-4} -8q_1^2t^{-2(1+p_1)}\xi^{-2(1+\hat{q_0})}K(\mathbf{p}, \hat{\mathbf{q}}), 
\end{equation}
\begin{align}\label{S4}
\nonumber \mathcal{I}_2 = & ~48 p_1^2p_2^2p_3^2t^{-6}\xi^{-6\hat{q_0}} -48q_0^2q_2^2q_3^2t^{-6p_1}\xi^{-6} \\ & 
 +24q_1^4t^{-2(1+2p_1)}\xi^{-2(2+\hat{q_0})}L_1(\mathbf{p}, \hat{\mathbf{q}}) 
 -24q_1^2t^{-2(2+p_1)}\xi^{-2(1+ 2\hat{q_0})}L_2(\mathbf{p}, \hat{\mathbf{q}}). 
\end{align}
Here we have employed the radial variable $\xi$ defined in Eq.~\eqref{M1}; moreover, to simplify the above invariants for the double-Kasner spacetime, we have made extensive use of the relations given in Appendix C. The expressions for $K$, $L_1$ and $L_2$ are in general  complicated and are given in Appendix D. 

Double-Kasner spacetime is a nonlinear superposition of two different Kasner spacetimes that are usually characterized as timelike and spacelike---cf. Eq.~\eqref{M2}, where we temporarily set $q_1=1$. We note that for the standard timelike Kasner spacetime,
\begin{equation}\label{S5} 
\mathcal{I}_1 = - 16 p_1p_2p_3t^{-4}, \quad  \mathcal{I}_2 = 48 p_1^2p_2^2p_3^2t^{-6}.
\end{equation}
Similarly, for the spacelike Kasner spacetime,
\begin{equation}\label{S6} 
\mathcal{I}_1 = -16\hat{q_0}\hat{q_2}\hat{q_3} \xi^{-4}, \quad  \mathcal{I}_2 = -48\hat{q_0}^2\hat{q_2}^2\hat{q_3}^2 \xi^{-6}.
\end{equation}
These invariants can be recognized as the components in the ``nonlinear superpositions" given in Eqs.~\eqref{S3} and~\eqref{S4}.

It follows from these curvature invariants that the double-Kasner spacetime has curvature singularities at $t = 0$ and $\xi = 0$, as expected. These can be generally characterized as spacelike and timelike, respectively. Moreover, it follows from Eqs.~\eqref{S3} and~\eqref{S4} that for $p_1 < 0$, there is an additional curvature singularity at $t = \infty$; similarly, for $\hat{q_0} < 0$, there is an additional curvature singularity at $\xi = \infty$. It is important to note that the physical spacetime domain under consideration in this paper is free of any singularities; in fact, curvature singularities occur at the boundaries of the admissible intervals of $t$ and $\xi$.

\section{Timelike Geodesics}
We consider a free test particle with four-velocity vector $u^\mu=dx^\mu/d\tau$ as before. The component of this vector along a Killing vector field is a constant of the motion; therefore, $g_2^2 dy/d\tau = C_2 $ and $g_3^2 dz/d\tau = C_3 $, where $C_2$ and $C_3$ are integrals of geodesic motion. Moreover, we find from $u_\mu u^\mu = -1$ that
\begin{equation}\label{N17}
e^{q_0x} \frac{dt}{d\tau} = \Big[ 1 + g_1^2 (\frac{dx}{d\tau})^2 + \frac{C_2^2}{g_2^2} + \frac{C_3^2}{g_3^2}\Big]^{1/2},
\end{equation}
where we have assumed that $t$ increases with $\tau$ along the geodesic world line. The geodesic equation for $x$ can be written as
\begin{equation}\label{N18}
\frac{d^2x}{d\tau^2} + 2 \frac{p_1}{t} \frac{dt}{d\tau} \frac{dx}{d\tau} + (q_1 + q_0) (\frac{dx}{d\tau})^2 = \frac{1}{g_1^2} \Big[-q_0 + (q_2-q_0) \frac{C_2^2}{g_2^2}+ (q_3-q_0) \frac{C_3^2}{g_3^2}\Big].
\end{equation}
Here we have used Eq.~\eqref{N17} as well as the connection coefficients for metric~\eqref{N1} given in Appendix B. It is useful to define a new function $W$ given by
\begin{equation}\label{N19}
W = t^{2p_1} e^{(q_1+q_0)x} \frac{dx}{d\tau}.
\end{equation}
Then the autonomous system of equations that we must numerically integrate is
\begin{equation}\label{N20}
 \frac{dt}{d\tau} = e^{-q_0x} \Big[ 1 + e^{2(q_1-q_0)x} \frac{W^2}{g_1^2} + \frac{C_2^2}{g_2^2} + \frac{C_3^2}{g_3^2}\Big]^{1/2},
\end{equation}
\begin{equation}\label{N21}
 \frac{dx}{d\tau} = e^{(q_1-q_0)x} \frac{W}{g_1^2},  \quad  \frac{dy}{d\tau} =  \frac{C_2}{g_2^2}, \quad \frac{dz}{d\tau} =  \frac{C_3}{g_3^2}, 
\end{equation}
\begin{equation}\label{N22}
\frac{dW}{d\tau} =  e^{-(q_1-q_0)x} \Big[-q_0 + (q_2-q_0) \frac{C_2^2}{g_2^2}+ (q_3-q_0) \frac{C_3^2}{g_3^2}\Big].
\end{equation}

Let us now imagine the fundamental observers in this spacetime with tetrad field $\lambda^\mu{}_{(\alpha)}$, where for $i = 1,2,3$,
\begin{equation}\label{N23}
\lambda^\mu{}_{(i)} = \frac{1}{g_i} \delta^\mu{}_{i}, \quad \lambda^\mu{}_{(0)} = e^{-q_0x}\delta^\mu{}_{0}. 
\end{equation}
It follows that the Lorentz factor of the free particle $\gamma$ and its velocity $\mathbf{v}$ as measured by the fundamental observers are given in terms of the system~\eqref{N20}--\eqref{N22} by
\begin{equation}\label{N24}
\gamma = e^{q_0x} \frac{dt}{d\tau}
\end{equation}
and
\begin{equation}\label{N25}
\gamma v_x = g_1 \frac{dx}{d\tau}, \quad  \gamma v_y = g_2 \frac{dy}{d\tau}, \quad \gamma v_z = g_3 \frac{dz}{d\tau}.
\end{equation}

If the $q$'s all vanish, $W = C_1$ is a constant and system~\eqref{N20}--\eqref{N22} reduces to the timelike geodesic equation for the standard Kasner spacetime, as expected. In that case, one of the $p$'s is negative (``collapse") and the other two are positive (``expansion"), resulting in an asymptotic double-jet pattern along the direction of collapse. However, the situation is generally different for the spatially inhomogeneous double-Kasner spacetime with nonvanishing $q$'s as demonstrated by the following numerical results.

\section{Numerical Results}

In integrating the system of equations~\eqref{N20}--\eqref{N22}, we must first fix the $p$'s, $q$'s and the specific momenta along the $y$ and $z$ directions given respectively by $C_2$ and $C_3$. Starting from initial conditions at $\tau = 0$ specified by $t(0) = t_0 =1$, $x(0) = x_0$, $y(0) = 0$,  $z(0) = 0$ and $W(0) = W_0$, we numerically integrate forward in proper time such that $t \to \infty$ and backward to $t = 0$. Let us note that in our dynamical system, the equations for $t(\tau)$, $x(\tau)$ and $W(\tau)$ are the primary coupled ordinary differential equations. This is due to translational invariance along the $y$ and $z$ directions; moreover, we can always choose the initial values of $y$ and $z$ at the origin of $(y,z)$ plane without any loss in generality. 

There are an infinite number of possible choices of $p$'s and $q$'s; once $\mathbf{p}$ is chosen,  the two possible  $\hat{\mathbf{q}}$ vectors can be algebraically determined, as discussed in Appendix C.  We then define $q_0 = \hat{q_0} \epsilon$, $q_1 = \epsilon$, $q_2 = \hat{q_2} \epsilon$ and $q_3 = \hat{q_3} \epsilon$. Here, $\epsilon$ is dimensionless, since $\epsilon = c t_0 q_1$ and $c=t_0=1$ in accordance with our Kasner conventions; in fact,   $ \epsilon \in [0, \infty)$ is such that for $\epsilon = 0$ we recover the standard timelike Kasner metric. Thus $\epsilon$ is a measure of how close the double-Kasner metric is to the original Kasner metric; moreover, $\epsilon^{-1}$ is a parameter that represents the extent of spatial inhomogeneities in the $x$ direction. It is important for the theory of cosmic double-jet configurations to recognize that for sufficiently small $\epsilon$, $0<\epsilon<<1$, general continuity arguments connect the double-Kasner geodesics to the Kasner geodesics over \emph{finite} intervals of proper time. The primary autonomous differential equations can be expressed as
\begin{equation}\label{N26}
 \frac{dt}{d\tau} = \xi^{- \hat{q_0}} \Big[ 1 + W^2 t^{-2 p_1}\xi^{-2\hat{q_0}}+ C_2^2 t^{-2 p_2}\xi^{-2\hat{q_2}} + C_3^2 t^{-2 p_3}\xi^{-2\hat{q_3}} \Big]^{1/2},
\end{equation}
\begin{equation}\label{N27}
 \frac{d\xi}{d\tau} =  \epsilon W t^{-2p_1} \xi^{-\hat{q_0}} ,  
\end{equation}
\begin{equation}\label{N28}
\frac{dW}{d\tau} = - \epsilon \xi^{\hat{q_0}-1} \Big[ \hat{q_0} + (\hat{q_0}-\hat{q_2})C_2^2 t^{-2 p_2}\xi^{-2\hat{q_2}} + (\hat{q_0}-\hat{q_3})C_3^2 t^{-2 p_3}\xi^{-2\hat{q_3}}\Big];
\end{equation}
furthermore, $dy/d\tau = C_2t^{-2 p_2}\xi^{-2\hat{q_2}}$ and $dz/d\tau = C_3t^{-2 p_3}\xi^{-2\hat{q_3}}$. For $W \ne 0$, the primary differential equations can also be written as
\begin{equation}\label{N28a}
\frac{dt}{d\xi} =  \frac{t^{p_1} \xi^{-\hat{q_0}}(W^2+ \mathcal{U})^{1/2}}{\epsilon W},  
\end{equation}
\begin{equation}\label{N28b}
\frac{d}{d\xi}W^2 =  -\frac{\partial \mathcal{U}(t,\xi)}{\partial \xi},  
\end{equation}
where $\mathcal{U}$ is given by
\begin{equation}\label{N28c}
\mathcal{U}(t,\xi)= t^{2 p_1}\xi^{2\hat{q_0}}(1+ C_2^2 t^{-2 p_2}\xi^{-2\hat{q_2}} + C_3^2 t^{-2 p_3}\xi^{-2\hat{q_3}}).
\end{equation}
Equation~\eqref{N28b} can be simplified in certain situations, thereby leading to the application of  the effective potential method in one-dimensional motion---see, for instance, Eqs.~\eqref{N29} and \eqref{N30} in Case 1 below.

The autonomous system of ordinary differential equations for timelike geodesics has singularities at $t=0$ and $\xi=0$. The family of solutions of this system depends continuously on proper time $\tau$ as well as the parameter $\epsilon$. These solutions converge uniformly to the solution with $\epsilon = 0$ on every finite time interval. That is, once a \emph{finite} interval of proper time is chosen, then for a sufficiently small $\epsilon$, the geodesics of double-Kasner spacetime behave as the geodesics of Kasner spacetime at least up to the end of this fixed time interval. 

We are interested in the main characteristics of peculiar motion in the double-Kasner spacetime; therefore, it is useful to mention for future reference that in terms of the radial coordinate $\xi$,
\begin{equation}\label{N25a}
\gamma v_x = Wt^{-p_1}\xi^{-\hat{q_0}}, \quad  \gamma v_y = C_2t^{-p_2}\xi^{-\hat{q_2}}, \quad \gamma v_z = C_3t^{-p_3}\xi^{-\hat{q_3}}.
\end{equation} 

A complete numerical analysis of timelike geodesic motion in the double-Kasner spacetime is beyond the scope of our investigation. To simplify matters, we therefore limit our attention to the five special cases listed in Eqs.~\eqref{N7}--\eqref{N11}, which will be treated in turn below. We hope that our approach captures the main features of peculiar motion that could be of interest in connection with cosmic jets.  

\subsection{Case 1: $\mathbf{p} = (0, 1, 0)$ and $\hat{\mathbf{q}} = (\frac{2}{3}, \frac{2}{3}, -\frac{1}{3})$}

The temporal dependence of the metric in this special case simply  consists of \emph{expansion} along the $y$ axis; moreover, the corresponding curvature invariants considered in Sec. IV essentially reduce to those of the spacelike form of the Kasner metric. In fact,  theses curvature scalars are given in Case 1 by~$q_1^4 \mathcal{I}_1$ and $q_1^6 \mathcal{I}_2$, where $\mathcal{I}_1$ and $\mathcal{I}_2$ are the spacelike Kasner invariants given in Eq.~\eqref{S6}.

Numerical experiments involving \emph{forward integration in time} indicate in this case the dominant attractive character of the timelike singularity at $\xi = 0$ corresponding to $x = - \infty$. In \emph{finite proper time}, all timelike geodesics appear to approach this singularity with a peculiar velocity that approaches the velocity of light. To see this analytically, let us find $dW/d\xi$ from Eqs.~\eqref{N27} and \eqref{N28}. In this case, the result is 
\begin{equation}\label{N29}
-W \frac{dW}{d\xi} =  \frac{2}{3} \xi^{\frac{1}{3}} + C_3^2 \xi.
\end{equation}
Numerical experiments indicate that $\xi = 0$ at $t = t_s > 1$, $\tau = \tau_s > 0$ and $W=W_s < 0$. Integrating Eq.~\eqref{N29}, we find 
\begin{equation}\label{N30}
W^2 + ( \xi^{\frac{4}{3}} + C_3^2 \xi^2)=W_s^2.
\end{equation}
It follows that if $W_0>0$ at $\xi(0)>0$, then a free test particle will move such that its $x$ component of motion monotonically increases and $W$ monotonically decreases until $x$, or equivalently $\xi$, reaches its maximum value where $W=0$. At this point, the test particle stops along the $x$ axis and changes direction such that subsequently $W<0$ and the particle falls toward the timelike singularity at $\xi=0$. It then follows from Eq.~\eqref{N25a} that
\begin{equation}\label{N31}
\gamma =  \xi^{- \frac{2}{3}} \Big(W_s^2 + \frac{C_2^2}{t^2}\Big)^{1/2}
\end{equation}
and as $\xi \to 0$, $\gamma \to \infty$ and
\begin{equation}\label{N32}
v_x \to -\frac{t_s|W_s|}{(C_2^2+t_s^2W_s^2)^{1/2}}, \quad v_y \to \frac{C_2}{(C_2^2+t_s^2W_s^2)^{1/2}}, \quad v_z \to 0,
\end{equation}
so that $v \to 1$.

Using Eq.~\eqref{N30}, it is possible to reduce the differential equations for the world line to quadratures in this case. It then follows from these results in the case of \emph{backward integration in time} that as $t \to 0$, $\xi$ approaches a finite nonzero number and 
\begin{equation}\label{N32a}
v_x \to 0, \quad v_y \to \frac{C_2}{|C_2|}, \quad v_z \to 0.
\end{equation}
Indeed, it is clear from Eq.~\eqref{N31} in this case that $\gamma \to \infty$ as $t \to 0$. It is important to recall here that the results for geodesic motion in the spatially inhomogeneous double-Kasner spacetime are not in general expected to be consistent with the general notion, developed for spatially homogeneous spacetimes in Sec. II, that peculiar velocities decrease to zero along expanding directions and increase up to the speed of light along contracting directions. However, throughout this section, our numerical results for \emph{backward integration in time} generally agree with those in Kasner spacetime, since $\xi$ generally approaches a finite nonzero value as $t \to 0$. To see the influence of spatial inhomogeneities on peculiar velocities, we henceforth concentrate on forward temporal integration. 

These analytic results illustrate a feature that is ubiquitous in the numerical results of \emph{forward integration} in the other cases as well, except for Case 4, where $\hat{q_0} < 0$. Therefore, let us assume that $\hat{q_0} > 0$ and we set $C_2 = C_3 = 0$ for the sake of simplicity. As we integrate forward from $\tau = 0$, we soon encounter the $\xi = 0$ singularity at $t = t_s > 1$, $\tau = \tau_s > 0$ and $W=W_s < 0$. Thus \emph{near} $\xi = 0$, under the square root in Eq.~\eqref{N26}, we can neglect the first term (i.e., unity) in comparison to the second; hence,
\begin{equation}\label{N33}
 \frac{dt}{d\tau} \approx - W t^{- p_1}\xi^{-2\hat{q_0}},
\end{equation}
\begin{equation}\label{N34}
 \frac{d\xi}{d\tau} =  \epsilon W t^{-2p_1} \xi^{-\hat{q_0}}.  
\end{equation}
Thus,
\begin{equation}\label{N35}
 \frac{d\xi}{dt} \approx  - \epsilon  t^{-p_1} \xi^{\hat{q_0}}.  
\end{equation}
Integrating this relation near the singularity results in
\begin{equation}\label{N36}
\xi^{1-\hat{q_0}} \approx  \epsilon (1-\hat{q_0}) t_s^{-p_1} (t_s - t).  
\end{equation}
Similarly consistent results can be obtained for $\tau$ and $W$ near the singularity. From Eq.~\eqref{N25a}, we find that as $\xi \to 0$, $\gamma$ diverges; indeed, near $\xi = 0$,
\begin{equation}\label{N37}
\gamma \approx (-W_s t_s^{-p_1}) \xi^{-\hat{q_0}},
\end{equation}
so that $v_x \to -1$, since $v_y = 0$ and $v_z = 0$.

For the sake of completeness, it is important to compare and contrast here the timelike motion of a test particle along the $x$ direction, discussed above, with that of a light ray. The equation of motion of the light ray follows from Eq.~\eqref{M2} and $ds^2=0$, namely, $d\xi/dt=\pm \epsilon t^{-p_1}\xi^{\hat{q_0}}$. We find that if the light ray is initially moving along the positive $x$ direction, then integrating the equation of motion with the upper sign implies that it will continue to do so and $x \to \infty$ as $t \to \infty$. On the other hand, if the light ray is initially moving in the negative $x$ direction, then integrating the equation of motion with the lower sign implies that it reaches the singularity at $x=-\infty$ in a finite interval of time, essentially as in the treatment presented above. 

\subsection{Case 2: $\mathbf{p} = (-\frac{1}{3}, \frac{2}{3}, \frac{2}{3})$ and $\hat{\mathbf{q}} = (\frac{3}{7}, \frac{6}{7}, -\frac{2}{7})$}
This case is particularly interesting as the direction of collapse coincides with the direction of spatial inhomogeneities in the double-Kasner spacetime. Thus for sufficiently small $\epsilon$, we expect that test particles following timelike geodesics would exhibit a double-jet configuration  parallel to the $x$ axis as in Kasner spacetime. This turns out to be true only if we limit our considerations to finite intervals of proper time. Beyond that, the singularity at $\xi = 0$ is expected to dominate the motion as discussed in Case 1. This duality has interesting consequences to which we now turn. 

Let us first recall that for $\epsilon = 0$, free test particles with nonzero peculiar velocity form an asymptotic Kasner double-jet pattern parallel to the $x$ axis as $t \to \infty$. To simplify matters, we assume that $C_2 = C_3 = 0$, so that we have geodesic motion only in the $x$ direction. Moreover, let us suppose that $\epsilon$, $0<\epsilon<<1$, is fixed. Then, a free test particle with a positive peculiar velocity characterized by $W_0>0$ is expected to move along the \emph{positive} $x$ direction and experience \emph{peculiar acceleration} just as in the Kasner spacetime. Numerical experiments show that this is indeed the case, as illustrated in figure 2, but that peculiar acceleration along the positive $x$ axis later changes to deceleration. The particle decelerates for a while until it stops ($\gamma = 1$) at a finite proper time and reverses course, moving along the negative $x$ direction until it eventually reaches the singularity  $x = - \infty ~(\xi = 0)$ after a long but finite proper time with its peculiar speed approaching unity $(\gamma \to \infty)$. To see how this comes about, we note that Eq.~\eqref{N28} can be written in this case as
\begin{equation}\label{N38}
\frac{dW}{d\tau} = - \frac{3}{7} \epsilon \xi^{- \frac{4}{7}}.
\end{equation}
Let us define a new parameter $\tau'$ along the geodesic world line such that $d\tau' = \xi^{-4/7} d\tau$, so that $\tau'$ increases with proper time $\tau$ along the path. It follows from Eq.~\eqref{N38} that $-dW/d\tau'$ is a positive constant that is much less than unity; therefore, $W$ monotonically decreases along the path. This means that even when $W$ is initially positive, $W_0>0$, it will eventually turn negative in finite proper time and then the general argument presented in Eqs.~\eqref{N33}--\eqref{N37} can be employed to show that the free test particle should stop at some time and then fall back toward the dominant singularity at $x = - \infty$. An example of this general behavior is given in Fig. 2 for $W_0 = 1$.  If $W_0$ is negative, the corresponding test particle simply moves along the negative $x$ direction and inevitably ends up at the $x = - \infty ~(\xi = 0)$ singularity with its peculiar velocity approaching the velocity of light, in accordance with the analysis contained in Eqs.~\eqref{N33}--\eqref{N37}.

When we allow $C_2$ and/or $C_3$ to be nonzero, the motion is in general more complicated due to the extra degrees of freedom; in fact, an interesting oscillatory behavior can in general occur that will be discussed in the last part of this section.

We conclude that even when spatial inhomogeneities are turned on, as in the double-Kasner spacetime, the Kasner double-jet pattern can survive over a finite time interval under favorable circumstances, but it is then significantly modified by the presence of spatial inhomogeneities. 

Finally, backward numerical integration in the case under consideration in this subsection, namely, $x_0 = 0, W_0 = 1$ and $C_2=C_3=0$, is consistent with the result that as $t \to 0$, $\gamma \to 1$ and $v_x \to 0$, while $v_y=v_z=0$ by assumption. That is, the motion is in this case confined to the $x$ direction, which expands as $t \to 0$ and, as expected, we find that $v_x \to 0$. The same result can be obtained from a straightforward mathematical analysis of our dynamical system near $t=0$. The geodesic equations of motion are simplified by considering the dominant terms for $t \to 0$; in fact, this can be done for all the subsequent cases discussed in this section. 

\begin{figure}
\includegraphics[scale=0.8,angle=0]{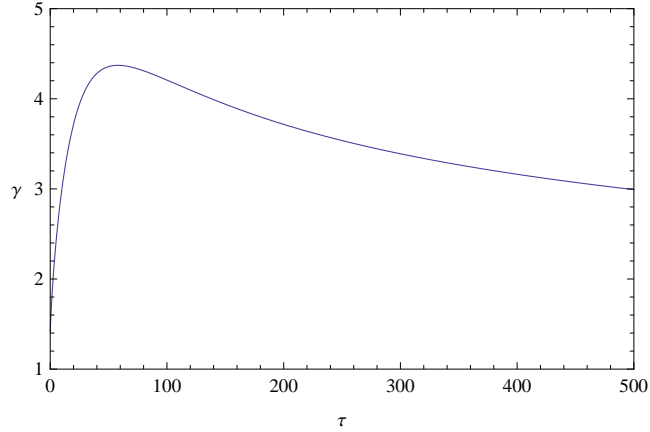}
\caption{Plot of the Lorentz factor $\gamma$ versus proper time $\tau$ for the peculiar motion of a free test particle along the $x$ direction in Case 2. The initial conditions at $\tau=0$ are $x_0=0$ and $W_0=1$; moreover, $\epsilon = 10^{-3}$ and $C_2=C_3=0$. Thus at $\tau=0$ we have $\gamma=\sqrt{2}\approx{1.4}$. The figure shows that the initial peculiar acceleration reaches a peak around $\tau=50$ and then turns to deceleration. Further numerical integration shows that the particle eventually stops at around $\tau=50000$, reverses course and accelerates toward the $\xi=0$ singularity with $\gamma \to \infty$.} \label{fig:2} 
\end{figure}

\subsection{Case 3: $\mathbf{p} = (\frac{2}{3}, -\frac{1}{3}, \frac{2}{3})$ and $\hat{\mathbf{q}} = (\frac{6}{7}, -\frac{2}{7}, \frac{3}{7})$}
In this case, the Kasner limit (for $\epsilon=0$) involves a double-jet pattern along the $y$ axis. Following our general approach, we set $\epsilon = 10^{-3}$ and numerically integrate the equations of motion forward in time with $x_0=0$, $W_0=0$, $C_2=\pm 1$ and $C_3=0$. As expected, the result of the integration is similar to the Kasner case over a certain initial time interval, but later the motion along the negative $x$ direction toward the timelike singularity at $x=-\infty$ takes over and $\gamma$ tends to infinity. That is, as $\xi \to 0$, $v_x^2+v_y^2 \to 1$, while $v_z=0$. The situation in the descent toward the singularity is essentially analogous to the analysis contained in Eqs.~\eqref{N33}--\eqref{N37} for $\hat{q_0} > 0$. The numerical results for $\gamma$ are presented in Fig. 3.

When we numerically integrate the equations of motion backward in time, we find that as $t \to 0$, $v_x \to 1, v_y \to 0$, while $v_z=0$. Thus $\gamma \to \infty$ and the peculiar velocity of a free test particle approaches the velocity of light toward the cosmological singularity at $t=0$. This conclusion is in agreement with a detailed theoretical analysis of the equations of motion in Case 3 near $t=0$.  

\begin{figure}
\includegraphics[scale=0.8,angle=0]{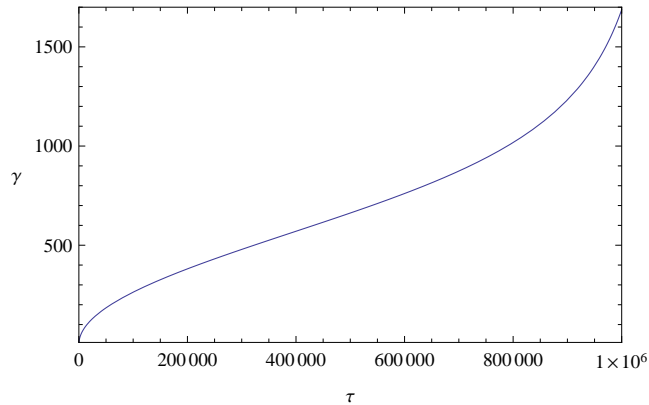}
\caption{Plot of the Lorentz factor $\gamma$ versus proper time $\tau$ for the peculiar motion of a free test particle in the $(x,y)$ plane in Case 3. Note that initially $\gamma=\sqrt{2}\approx{1.4}$ at $\xi=1$, but in time $\gamma \to \infty$ as the $\xi=0$ singularity is approached. The initial conditions at $\tau=0$ are $x_0=0$ and $W_0=0$; moreover, $\epsilon = 10^{-3}$, $C_2= \pm 1$ and $C_3=0$.}   \label{fig:3} 
\end{figure}

\subsection{Case 4: $\mathbf{p} = (\frac{3}{7}, \frac{6}{7}, -\frac{2}{7})$ and $\hat{\mathbf{q}} = (-\frac{1}{3}, \frac{2}{3}, \frac{2}{3})$}
This case is qualitatively different from the other cases as $\hat{q_0} < 0$, which means that an analysis similar to that contained in Eqs.~\eqref{N33}--\eqref{N37} for forward integration does not apply here; that is, forward integration in time is not dominated by the singularity at $\xi=0$. The Kasner limit (for $\epsilon=0$) involves a double-jet pattern along the $z$ axis, which we expect can persist here for a finite interval of time when $\epsilon$ is sufficiently small.  Our numerical experiments indicate similar qualitative behavior for $\epsilon \sim 10^{-3}$ as for $\epsilon \sim 1$. Therefore, we set $\epsilon = 1$ and integrate the equations of motion forward in time with $x_0=0$, $W_0=0$, $C_2=0$ and $C_3=\pm 1$. The initial peculiar acceleration later turns to deceleration until the peculiar velocity of a free test particle approaches a constant \emph{terminal velocity} vector whose magnitude is always less than the speed of light. The corresponding behavior of $\gamma$ is presented in Fig. 4. If the initial value of $W$ is negative, $W_0<0$, the free test particle's initial movement along the negative $x$ direction comes to a halt after a while, the particle reverses course and, as before, approaches $\xi = \infty$ with a terminal peculiar speed that is less than unity. The existence of the turning point in this case appears to imply that there is a barrier blocking the particle's access to the $\xi = 0$ singularity. For large $\xi$, $\xi \to \infty$, the asymptotic behavior of the equations of motion can be worked out in this case, and we find that  $t \sim \xi^{7/3}$, $\tau \sim \xi^2$ and $W \sim \xi^{2/3}$.

Let us recall here that when $\hat{q_0} > 0$, the only timelike singularity is the one at
$\xi = 0$. This strongly attracts free test particles, which can reach $\xi = 0$ in finite values of
proper time. When $\hat{q_0} < 0$, however, there is an additional timelike singularity
at $\xi = \infty$. Indeed, our numerical results in this case seem to indicate that the new singularity at $\xi = \infty$ is dominant for $\hat{q_0} < 0$, while the $\xi=0$ singularity is inactive. Moreover, the $\xi=\infty$ singularity seems to be somehow weaker in terms of its attractive character: free test
particles approach $\xi = \infty$ with proper times that go to infinity and terminal peculiar speeds that are less than the speed of light. 

This case brings out a certain generic behavior of timelike geodesics for $\hat{q_0} < 0$. To illustrate this point, let us set $C_2=C_3=0$ for the sake of simplicity and assume that as $\xi \to \infty$, the asymptotic behaviors of $t$, $\tau$ and $W$ are given by
\begin{equation}\label{N39}
t \sim \xi^{\kappa_1}, \quad  \tau \sim \xi^{\kappa_2}, \quad W \sim \xi^{\kappa_3}.
\end{equation}
A detailed analysis of the autonomous system~\eqref{N26}--\eqref{N28} reveals that this assumption is valid with $\kappa_1$, $\kappa_2$ and $\kappa_3$ all positive and given by
\begin{equation}\label{N40}
\kappa_1 = \frac{1-\hat{q_0}}{1-p_1}, \quad  \kappa_2 = 1+ p_1 \kappa_1, \quad \kappa_3=\hat{q_0} + p_1 \kappa_1.
\end{equation}
Moreover, it is possible to show that the motion along the $x$ direction reaches a terminal peculiar speed as $\xi \to \infty$ with Lorentz factor $\gamma_{\infty}$,
\begin{equation}\label{N41}
\gamma_{\infty} = \Big(1- \frac{\hat{q_0}}{\kappa_3}\Big)^{1/2}.
\end{equation}

The result of the integration of the equations of motion backward in time is that as $t \to 0$, $v_x \to -1$, $v_y=0$ and $v_z \to 0$. This means that the peculiar velocity of a free test particle approaches the velocity of light and $\gamma \to \infty$ as $t \to 0$. This result is consistent with a detailed analytic treatment of the equations of motion in this case near $t=0$. 

\begin{figure}
\includegraphics[scale=0.8,angle=0]{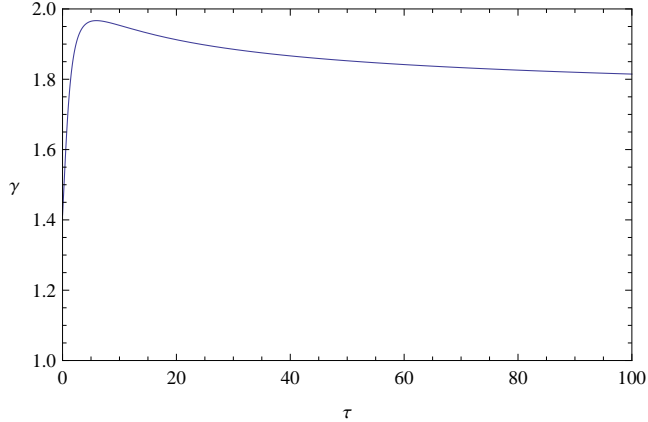}
\caption{Plot of the Lorentz factor $\gamma$ versus proper time $\tau$ in Case 4. The initial conditions at $\tau=0$ are $x_0=0$ and $W_0=0$; moreover, $\epsilon = 1$, $C_2= 0$ and $C_3= \pm 1$. Note that $\gamma$ is initially $\sqrt{2}\approx{1.4}$ and approaches a finite terminal value as $\tau \to \infty$.}  \label{fig:4} 
\end{figure}

\subsection{Case 5: $\mathbf{p} = (\frac{6}{7}, -\frac{2}{7}, \frac{3}{7})$ and $\hat{\mathbf{q}} = (\frac{2}{3}, -\frac{1}{3}, \frac{2}{3})$}
It is interesting to point out that if in this case we switch $\mathbf{p}$ and $\hat{\mathbf{q}}$, we get Case 3. There is a similar connection between Cases 2 and 4. Case 1 stands alone, however, since switching $\mathbf{p}$ and $\hat{\mathbf{q}}$ in this case leads to Kasner spacetime.

Let us note that the Kasner limit ($\epsilon=0$) in this case involves a double-jet pattern along the $y$ axis, as in Case 3. Therefore, just as in Case 3,  we set $\epsilon = 10^{-3}$ and integrate the equations of motion forward in time with $x_0=0$, $W_0=0$, $C_2=\pm 1$ and $C_3=0$. Though the details are different, the end result of forward integration is qualitatively the same as in Case 3; in fact, this is also the case when we integrate the equations of motion backward in time toward $t = 0$.

\subsection{Concluding Remarks}

We conclude this section with some general observations. Starting from the present cosmic epoch $t=t_0=1$ and integrating the geodesic equations \emph{backward} to $t=0$, we have found that the behavior of double-Kasner peculiar velocities near the cosmological singularity is essentially the same as in Kasner spacetime. Therefore, we turn to \emph{forward} integration ($t: 1\to \infty$). Here, beyond the initial Kasner double-jet configuration, our limited numerical results for \emph{late-time} peculiar motion in the double-Kasner spacetime depend significantly on whether $\hat{q_0}$ is positive or negative. For  $\hat{q_0}>0$, there seems to be a strong source of gravitational attraction at the timelike curvature singularity $\xi=0$. We note that for the double-Kasner metric, $g_{tt}= -\xi^{2 \hat{q_0}}$, so that $g_{tt} \to 0$ as $\xi \to 0$ for $\hat{q_0}>0$. In this limit, furthermore, the peculiar velocity approaches the velocity of light. On the other hand, for $\hat{q_0}<0$, $g_{tt} \to 0$ as $\xi \to \infty$; in this case, $\xi=\infty$ also happens to be a timelike curvature singularity according to the results of Sec. IV. Qualitatively, for $\hat{q_0}<0$, the $\xi=0$ singularity is somehow inactive and the dominant attractive influence is exerted by the singularity at $\xi=\infty$, which results in uniform peculiar motion at late times ($\tau \to \infty$ and $\xi \to \infty$).

To find an explanation for this type of late-time behavior, we must turn to the other half of the double-Kasner geometry. Let us recall here that the double-Kasner solution is a certain nonlinear superposition of the standard timelike Kasner solution~\eqref{eq:6} and the spacelike Kasner solution given by
\begin{equation}\label{N42}
ds^2=-\xi^{2\hat{q_0}}dt^2+  d\xi^2+\xi^{2\hat{q_2}} dy^2+\xi^{2\hat{q_3}}dz^2.
\end{equation}
This is a static Ricci-flat solution of general relativity with commuting Killing vector fields $\partial_t$, $\partial_y$ and $\partial_z$. The admissibility conditions restrict the radial coordinate $\xi$ such that $\xi \in (0,\infty)$; moreover, there exists a curvature singularity at $\xi=0$. The motion of a free test particle in this spacetime is such that the components of the four-velocity vector of the particle $u^{\mu}$ along the Killing vectors are constants of the motion; that is, $u \cdot \partial_t=-E_0$, $u \cdot \partial_y=E_2$ and $u \cdot \partial_z=E_3$. Here $E_0>0$ is the constant specific energy of the particle, while, as before, $E_2$ and $E_3$ are constant specific momenta. The geodesic equation of motion for the radial coordinate $\xi$ of the test particle then follows from $u \cdot u = -1$; that is, 
\begin{equation}\label{N43}
\Big(\frac{d\xi}{d\tau}\Big)^2 + \mathcal{V}(\xi) = -1,
\end{equation}
where $\mathcal{V}$ is the effective potential given by 
\begin{equation}\label{N44}
 \mathcal{V}(\xi) = -\frac{E_0^2}{\xi^{2\hat{q_0}}} + \frac{E_2^2}{\xi^{2\hat{q_2}}} +\frac{E_3^2}{\xi^{2\hat{q_3}}}.
\end{equation}

For $\hat{q_0}>0$, the effective potential is such that the motion can be either confined within the interval $(0,\xi_{\max}]$ and may then be described as a ``fall" toward $\xi=0$, or it can be oscillatory within the interval $[\xi_{\min},\xi_{\max}]$. Here $\xi_{\min}$ and $\xi_{\max}$ are turning points such that $0<\xi_{\min}<\xi_{\max}$, or the turning points could coincide, in which case $\xi$ would be fixed at the minimum of the effective potential. Only the fall toward the curvature singularity $\xi=0$ is available for $E_2=E_3=0$. On the other hand, for  $\hat{q_0}<0$, the effective potential for $\xi: 0 \to \infty$ is monotonically decreasing from $+\infty$ to $-\infty$, so that radial motion in $\xi$ is confined within the interval $[\xi_{\min}, \infty)$ and may be described as ``escape" to infinity. These possibilities for radial motion may be compared and contrasted with geodesic motion in exterior Schwarzschild spacetime. Furthermore, we note that $\partial_t \cdot \partial_t = - \xi^{2\hat{q_0}}$, so that the timelike Killing vector becomes null at the endpoints that we have discussed here: the curvature singularity $\xi=0$ for $\hat{q_0}>0$ and $\xi=\infty$ for $\hat{q_0}<0$.

It remains to discuss the possibility of oscillatory motion within the interval $[\xi_{\min},\xi_{\max}]$. Let us assume, for instance, that  $\hat{q_2}>\hat{q_0}>0$ and $\hat{q_3}<0$; then, oscillatory motion is possible for $E_2 \ne 0$. Moreover, we expect time-dependent oscillatory peculiar motion in the more general context of the double-Kasner spacetime. To see an example of this behavior, we return to Case 2 above and assume that  $\epsilon = 1$, $x_0=-1$ and $W_0 =0$; moreover, we set $C_2= 1$ and $C_3=2$. The time-dependent oscillatory character of the Lorentz factor for peculiar motion in this case is illustrated in Fig. 5.  We intuitively expect the turning points to be time-dependent in this case; in fact, in time $\xi_{\min}$  approaches the singularity at $\xi=0$ in the case depicted in Fig. 5.

\begin{figure}
\includegraphics[scale=0.8,angle=0]{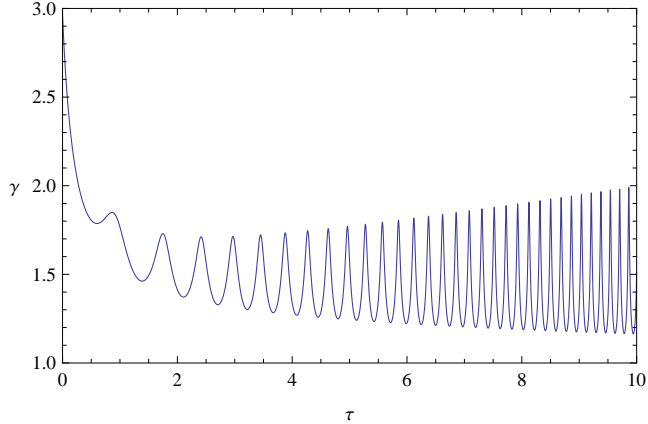}
\caption{Plot of the oscillatory Lorentz factor $\gamma$ versus proper time $\tau$ in Case 2. The initial conditions at $\tau=0$ are $x_0=-1$ and $W_0=0$; moreover, $\epsilon = 1$, $C_2= 1$ and $C_3= 2$. Note that the Lorentz factor is initially $\gamma \approx{2.97}$. The components of the peculiar velocity are all oscillatory with increasing amplitudes for $v_x$ and $v_y$ and decreasing amplitude for $v_z$. In fact, $v_z$ tends to zero as $\xi_{\min}$ approaches zero.} \label{fig:5} 
\end{figure}

\section{discussion}
From the spatially homogeneous examples discussed in Sec. II of this paper, one may draw some tentative conclusions (compare, in particular,  Fig.~\ref{fig:1}). As a general remark, the occurrence of jets requires peculiar acceleration; therefore, jets occur in more special situations compared to peculiar acceleration itself, which is consequently a more general phenomenon.  Restricting attention first to peculiar acceleration, we need some basic ingredients to describe it, and to characterize its properties. To begin with, we must define an appropriate background field of observers. This is because to define peculiar acceleration as such, we need some reference background, since only relative accelerations are possible when gravitational interactions are involved. In typical physical situations, there is, in most cases, a preferred observer family. For example, in spatially homogeneous (non-tilted) models, such as Kasner spacetime, the preferred family corresponds to the normals of the homogeneous hypersurfaces (see Sec. II). 

As for the observed objects, it is natural to consider them as members of a test field of geodesics. It is the peculiar velocity and acceleration of this test field with respect to the background observers that is the subject of investigation here. The relative velocity can be identified with the peculiar velocity in the astrophysical sense. Therefore, let $v$ be the peculiar velocity of a free test particle relative to the background observer family.  
Then our findings indicate that along any expanding axis $v \to  0$  as expansion tends to infinity, while along any contracting axis $v \to \pm c$ as contraction tends to zero. 
Relating this behavior to the background flow, it appears that
if in an expanding direction the flow is stable (corresponding to attraction) then there is no acceleration. Whereas, in a contracting direction, if the flow is unstable (repulsion), then there is acceleration.

In the gravitational collapse of an astrophysical object such as a star, the peculiar motion of free test particles would be referred to the collapsing medium, so that at first sight peculiar acceleration might appear as a simple kinematic effect of relative motion. However, the limiting situation of peculiar acceleration to the speed of light---emphasized in [1] and [2]---is observer-independent and seems to us to indicate true transfer of energy from the gravitational field to the free test particles in analogy with the electromagnetic acceleration of charged particles. Indeed, this gravitational transfer of energy is expected to be a general feature of expanding or contracting geodesic congruences in spacetimes that have no timelike Killing vector field; in these gravitational fields, energy gain is associated with collapse and energy loss is associated with expansion. The energy loss in the context of an expanding FLRW model has been discussed by Harrison~\cite{H}. We have emphasized energy gain in connection with gravitational collapse. However, in realistic collapse scenarios of astrophysical interest involving jets and cosmic rays, it remains to see if the gravitational energy gain associated with peculiar acceleration is in fact significant. This important problem is beyond the scope of our investigation. Instead, we have concentrated in this work on the simpler problem of the behavior of peculiar motions in the spatially inhomogeneous and anisotropic double-Kasner spacetime.

The double-Kasner metric reduces to the standard timelike Kasner metric when the inhomogeneity  parameter $\epsilon$ is zero. Thus when $\epsilon$ is sufficiently small, $0<\epsilon<<1$, we expect on the basis of continuity arguments that the Kasner double-jet pattern of peculiar motions would persist over a finite time interval. This general behavior is indeed confirmed by our numerical work. At late times, however, a significant feature of peculiar motions is that the jets tend to move toward either $x=-\infty~(\xi=0)$ or $x=\infty~(\xi=\infty)$ depending on whether $\hat{q_0}>0$ or $\hat{q_0}<0$, respectively. For $\hat{q_0}>0$, it follows from forward integration in proper time that at late times the dominant aspect of the peculiar motions in the double-Kasner spacetime is the strong gravitational attraction toward the timelike curvature singularity at $x=-\infty~(\xi=0)$ with a peculiar velocity approaching the velocity of light. On the other hand, for $\hat{q_0}<0$, we find jets moving toward another timelike curvature singularity at $x=\infty~(\xi=\infty)$ with uniform terminal speeds that stay well below the speed of light. We have shown that these aspects are associated with the other (spacelike) Kasner geometry that is part of the double-Kasner gravitational field. In these considerations the $x$ axis is distinguished from the other spatial axes in double-Kasner spacetime, since all spatial inhomogeneities occur along the $x$ direction. 

Finally, it is interesting to note that recent measurements of large-scale peculiar velocities of clusters of galaxies have been interpreted in terms of an anomalously rapid bulk flow in a common direction (``dark flow")---see~\cite{K1, K2, DC, MA} and the references cited therein. If confirmed, this dark flow would indicate a significant departure from the presumed large-scale spatial homogeneity and isotropy of the standard FLRW cosmology.

\begin{acknowledgments}
BM is grateful to Dejan Stojkovic for helpful discussions. CC was supported in part by the NSF grant DMS 0604331.
\end{acknowledgments}

\appendix{}
\section{peculiar velocities in FLRW models}\label{app:A}
The spatial isotropy and homogeneity of the FLRW universe imply that in the standard $(t, \chi, \theta, \phi)$ coordinates about any point in space as origin of spherical polar coordinates $(r,\theta, \phi)$---with $r = R_0 \chi, R_0 \sin\chi$ or $R_0 \sinh\chi$, respectively, for flat, closed or open models---the geodesics issuing from this spatial origin are purely radial. Here $R_0$ is the radius of curvature at the present epoch. Therefore, the geodesic equations are such that $\theta$ and $\phi$ are fixed while
\begin{align}\label{eq:A1}
\frac{dt}{d\tau} = \Big(1 + \frac{D^2}{a^2}\Big)^{1/2},
\end{align}
\begin{align}\label{eq:A2}
R_0 \frac{d\chi}{d\tau} = \frac{D}{a^2},
\end{align}
where $D$ is a dimensionless constant of integration and $a$, $a(t_0) = 1$, is the scale factor. Thus with respect to the natural tetrad frame of the fundamental observer at the spatial origin, the nonzero components of $u_{(\alpha)}$ are given by $u_{(t)} = -dt/d\tau$ and $u_{(\chi)} = D/a$. It follows that 
\begin{align}\label{eq:A3}
\gamma = \Big(1 + \frac{D^2}{a^2}\Big)^{1/2},\quad \gamma \mathbf{v} = \Big(\frac{D}{a}, 0, 0\Big).
\end{align}
Hence as $a\to 0$, $v^2 \to 1$. In particular, this proves that in general $\mathcal{P}(t) \propto a(t)^{-1}$, which is the law of variation of peculiar velocities in standard cosmological models~\cite{JP}.

\section{Christoffel Symbols}\label{app:B}
The nonzero components of the connection---modulo their symmetry $\Gamma^{\alpha}_{\beta\gamma} =  \Gamma^{\alpha}_{\gamma \beta}$---for metric~\eqref{N1} are given by $\Gamma^t_{tx} = q_0$,
\begin{equation}\label{eq:B1}
\Gamma^t_{xx} = \frac{p_1}{t}g_1^2e^{-2q_0x},
\end{equation}
\begin{equation}\label{eq:B2}
\Gamma^t_{yy} = \frac{p_2}{t}g_2^2e^{-2q_0x},
\end{equation}
\begin{equation}\label{eq:B3}
\Gamma^t_{zz} = \frac{p_3}{t}g_3^2e^{-2q_0x}.
\end{equation}
Moreover, we have
\begin{equation}\label{eq:B4}
\Gamma^x_{tt} = q_0\frac{e^{2 q_0 x}}{g_1^2}, \quad  \Gamma^x_{tx} = \frac{p_1}{t},   
\end{equation}
\begin{align}\label{eq:B5}
\Gamma^x_{xx} = q_1, \quad \Gamma^x_{yy} = -q_2(\frac{g_2}{g_1})^2, \quad   \Gamma^x_{zz} = -q_3(\frac{g_3}{g_1})^2.
\end{align}
Finally, we find that
\begin{align}\label{eq:B6}
\Gamma^y_{ty} = \frac{p_2}{t}, \quad   \Gamma^y_{xy} = q_2
\end{align}
and
\begin{align}\label{eq:B7}
\Gamma^z_{tz} = \frac{p_3}{t}, \quad   \Gamma^z_{xz} = q_3.
\end{align}

\section{Relations Involving $\mathbf{p}$ and $\hat{\mathbf{q}}$}\label{app:C}

We start with the defining properties of $\mathbf{p}$ and $\hat{\mathbf{q}}$, namely, 
\begin{equation}\label{eq:C1}
\sum_i p_i = 1, \quad  \hat{q_0}+\hat{q_2}+\hat{q_3} = 1,
\end{equation}
\begin{equation}\label{eq:C2}
\sum_i p_i^2 = 1, \quad \hat{q_0}^2+\hat{q_2}^2+\hat{q_3}^2 = 1.
\end{equation}
For simplicity, the following consequences of Eqs.~\eqref{eq:C1} and~\eqref{eq:C2} are stated only for $\mathbf{p}$ with the understanding that corresponding relations hold for $\hat{\mathbf{q}}$ as well.

It follows from squaring Eq.~\eqref{eq:C1} and then using Eq.~\eqref{eq:C2} that
\begin{equation}\label{eq:C3}
p_1p_2+p_1p_3+p_2p_3 = 0.
\end{equation}
Next, from Eqs.~\eqref{eq:C1} and~\eqref{eq:C3} we find that for each $ i = 1,2,3,$
\begin{equation}\label{eq:C4}
p_i^2 (p_i - 1) = p_1p_2p_3.
\end{equation}
Squaring Eq.~\eqref{eq:C3} and then using Eq.~\eqref{eq:C1} result in 
\begin{equation}\label{eq:C5}
p_1^2p_2^2+p_1^2p_3^2+p_2^2p_3^2 = -2p_1p_2p_3.
\end{equation}
Equations~\eqref{eq:C4} and~\eqref{eq:C2} imply that 
\begin{equation}\label{eq:C6}
\sum_i p_i^3 = 1+3p_1p_2p_3.
\end{equation}
Similarly, it follows from  Eq.~\eqref{eq:C5} and the square of  Eq.~\eqref{eq:C2} that
\begin{equation}\label{eq:C7}
\sum_i p_i^4 = 1+4p_1p_2p_3.
\end{equation}
Multiplying Eqs.~\eqref{eq:C3} and~\eqref{eq:C5} together, and then using Eqs.~\eqref{eq:C1} and~\eqref{eq:C6}, we get
\begin{equation}\label{eq:C8}
p_1^3p_2^3+p_1^3p_3^3+p_2^3p_3^3 =  3p_1^2p_2^2p_3^2;
\end{equation}
alternatively, one can cube Eq.~\eqref{eq:C3}. We find from Eqs.~\eqref{eq:C3}, \eqref{eq:C5} and~\eqref{eq:C8}, via division by powers of $p_1p_2p_3 \ne 0$, that
\begin{equation}\label{eq:C9}
\sum_i\frac{1}{p_i} = 0, \quad \sum_i\frac{1}{p_i^2} = - \frac{2}{p_1p_2p_3}, \quad \sum_i\frac{1}{p_i^3} = \frac{3}{p_1p_2p_3}.
\end{equation}
Next, multiplying Eqs.~\eqref{eq:C2} and~\eqref{eq:C6} together, and then using Eqs.~\eqref{eq:C1}, \eqref{eq:C3} and~\eqref{eq:C5}, we find
\begin{equation}\label{eq:C10}
\sum_i p_i^5 = 1 + 5p_1p_2p_3.
\end{equation}
Similarly, multiplying Eqs.~\eqref{eq:C2} and~\eqref{eq:C7} together, and then using Eqs.~\eqref{eq:C2} and~\eqref{eq:C5}, we find
\begin{equation}\label{eq:C11}
\sum_i p_i^6 = 1 + 6p_1p_2p_3 + 3p_1^2p_2^2p_3^2;
\end{equation}
alternatively, we can square Eq.~\eqref{eq:C6} and then use Eq.~\eqref{eq:C8}.

Finally, let us note that the connection between $\mathbf{p}$ and $\hat{\mathbf{q}}$, given by Eq.~\eqref{N5}, can be written as 
\begin{equation}\label{eq:C12}
\mathbf{p} \cdot \hat{\mathbf{q}} = p_1(1-\hat{q_0}) + \hat{q_0}.
\end{equation}
This relation is \emph{invariant} under the exchange of $\mathbf{p}$ and $\hat{\mathbf{q}}$, since the right-hand side of Eq.~\eqref{eq:C12} can be written as $ \hat{q_0}(1-p_1) + p_1$. The general reciprocity between $\mathbf{p}$ and $\hat{\mathbf{q}}$ is noteworthy. Excluding the flat spacetime case with $\mathbf{p} = \hat{\mathbf{q}} = (1,0,0)$ given in Eq.~\eqref{N6}, let us introduce
\begin{equation}\label{eq:C13}
\alpha = \frac{p_2-p_1}{p_2+p_3},  \quad  \beta = \frac{p_3-p_1}{p_2+p_3},
\end{equation}
so that Eq.~\eqref{eq:C12} can be written as $ \hat{q_0} = \alpha \hat{q_2} + \beta \hat{q_3}$. It is interesting to note that
\begin{equation}\label{eq:C14}
(\alpha - \beta)^2 + 2(\alpha + \beta) = 3. 
\end{equation}
Once one member of the pair  $(\mathbf{p}, \hat{\mathbf{q}})$ is chosen, the other can be algebraically determined. Suppose, for instance, that $\mathbf{p}$ has been fixed; then, one possible $\hat{\mathbf{q}}$ is given by
\begin{equation}\label{eq:C15}
\hat{q_0}' = \frac{1}{2 \Delta}(2 \alpha \beta + \alpha -3 \beta +3),
\end{equation}
\begin{equation}\label{eq:C16}
\hat{q_2}' = \frac{1}{2 \Delta}( - \alpha^2 + \alpha \beta -\alpha +6),
\end{equation}
\begin{equation}\label{eq:C17}
\hat{q_3}' = \frac{1}{2 \Delta}(\alpha^2 - \alpha \beta -2 \alpha + \beta -1),
\end{equation}
where $\Delta$ can be expressed as 
\begin{equation}\label{eq:C18}
\Delta =  \alpha \beta - \alpha - \beta + 4.
\end{equation}
There is, however, a second possible $\hat{\mathbf{q}}$, which is given by
\begin{equation}\label{eq:C19}
\hat{q_0}'' = \frac{1}{2 \Delta}(2 \alpha \beta - 3 \alpha + \beta +3),
\end{equation}
\begin{equation}\label{eq:C20}
\hat{q_2}'' = \frac{1}{2 \Delta}(\beta^2 - \alpha \beta +  \alpha - 2 \beta -1),
\end{equation}
\begin{equation}\label{eq:C21}
\hat{q_3}'' = \frac{1}{2 \Delta}( - \beta^2 + \alpha \beta -\beta +6).
\end{equation}
We note that under the interchange of $p_2$ with $p_3$, or equivalently, of $\alpha$ with $\beta$,  the first possible $\hat{\mathbf{q}}=(\hat{q_0}, \hat{q_2}, \hat{q_3})$ transforms into the second, but with $\hat{q_2}$ and $\hat{q_3}$ interchanged.

Let us now turn to a useful Kasner index parameterization due to Lifshitz and Khalatnikov---see~\cite{LK,BC} and the references cited therein. It is interesting to extend this parameterization to  double-Kasner spacetime. Introducing $\Sigma(w)$, 
\begin{equation}\label{eq:C22}
\Sigma (w) = 1 + w + w^2, 
\end{equation}
we define parameters $U$ and $V$ as follows
\begin{equation}\label{eq:C23}
\mathbf{p}=\frac{1}{\Sigma(U)} \Big(-U,\quad 1+U, \quad U(1+U)\Big),
\end{equation}
\begin{equation}\label{eq:C24}
\hat{\mathbf{q}}=\frac{1}{\Sigma(V)} \Big(-V,\quad 1+V, \quad V(1+V)\Big).
\end{equation}
Using this parameterization, Eqs.~\eqref{eq:C1} and~\eqref{eq:C2} are automatically satisfied. On the other hand, Eq.~\eqref{eq:C12}, the relation between $\mathbf{p}$ and $\hat{\mathbf{q}}$, reduces to 
\begin{equation}\label{eq:C25}
(1+2V+UV)(1+2U+UV)=0.
\end{equation}
For a given $\mathbf{p}$, $U$ is fixed and there are two possible solutions for $V$, namely,
\begin{equation}\label{eq:C26}
V'=-\frac{1}{2+U}, \quad V''=-\frac{1+2U}{U},
\end{equation}
corresponding to $\hat{\mathbf{q}}'$ and $\hat{\mathbf{q}}''$, respectively. That is, one can directly verify that the substitution of $V'$ for $V$ in Eq.~\eqref{eq:C24} leads to Eqs.~\eqref{eq:C15}--\eqref{eq:C17}, where
\begin{equation}\label{eq:C27}
\alpha=\frac{1+2U}{(1+U)^2}, \quad \beta=\frac{U(2+U)}{(1+U)^2}.
\end{equation}
Similarly, the substitution of $V''$ for $V$ in Eq.~\eqref{eq:C24} leads to Eqs.~\eqref{eq:C19}--\eqref{eq:C21}. Explicitly, we have
\begin{equation}\label{eq:C28}
\Sigma_1~ \hat{\mathbf{q}}'= \Big(2+U,\quad (2+U)(1+U), \quad -1-U\Big),
\end{equation}
\begin{equation}\label{eq:C29}
\Sigma_2~ \hat{\mathbf{q}}''= \Big(U(1+2U),\quad -U(1+U), \quad (1+U)(1+2U)\Big),
\end{equation}
where
\begin{equation}\label{eq:C30}
\Sigma_1= 3+3U+U^2, \quad  \Sigma_2= 1+3U+3U^2.
\end{equation}
The same interchange property noted above is recovered in this parameterization when $U$ is replaced by $1/U$.

Let us note here, for the sake of concreteness, that with our choice of Kasner $\mathbf{p}$ exponents, namely, $p_1 \le p_2 \le p_3$, $U \in [1,\infty)$. In general, there are six possible permutations of the Kasner  $\hat{\mathbf{q}}$ exponents; for instance, $V \to -1-V$ merely interchanges $\hat{q_0}$ and $ \hat{q_2}$ in Eq.~\eqref{eq:C24}, while leaving $\hat{q_3}$ invariant. Such permutations divide the real $V$ axis into six equivalent intervals with endpoints that are given by the set $\{-\infty, -2, -1, -\frac{1}{2}, 0, 1, \infty\}$---see Fig. 4 of Ref.~\cite{BC}.

\section{$K$, $L_1$ and $L_2$}\label{app:D}

The purpose of this appendix is to give the expression for $K$, which appears in the Kretschmann scalar $\mathcal{I}_1$, as well as the expressions for  $L_1$ and $L_2$, which appear in the expression for $\mathcal{I}_2$ in Sec. IV.

The curvature tensor can be computed using the double-Kasner metric~\eqref{N1} and the corresponding expressions for curvature invariants can be simplified using the relations involving $\mathbf{p}$, given explicitly in Appendix C, and similar ones involving $\hat{\mathbf{q}}$, but \emph{not} the connection between $\mathbf{p}$ and $\hat{\mathbf{q}}$. The results for $K$, $L_1$ and $L_2$ are then given by:
\begin{align}\label{eq:D1}
\nonumber K(\mathbf{p}, \hat{\mathbf{q}}) = &~p_2^2(\hat{q_0}^2 - \hat{q_0}\hat{q_2}+\hat{q_2}^2) + p_2(- \hat{q_0}\hat{q_2}+ p_3 \hat{q_2}\hat{q_3}) + p_1^2(1-\hat{q_0}) \\ &
\nonumber -p_1\hat{q_0}^2  -  p_1p_2\hat{q_2}(1+\hat{q_2})-p_1p_3\hat{q_3}(1+\hat{q_3})+p_1\hat{q_0}(1+2p_2\hat{q_2}+2p_3\hat{q_3}) \\ &
-p_3\hat{q_0}\hat{q_3}+p_3^2(\hat{q_0}^2 - \hat{q_0}\hat{q_3}+\hat{q_3}^2),
\end{align}
\begin{align}\label{eq:D2}
\nonumber L_1(\mathbf{p}, \hat{\mathbf{q}}) = &~p_1^2\hat{q_0}(\hat{q_2}^3 + \hat{q_3}^3)-p_1\hat{q_2}^2 \hat{q_3}^2+ p_2^2 \hat{q_0}\hat{q_2}^2(2- \hat{q_2}+ 2\hat{q_3}) + p_3^2 \hat{q_0}\hat{q_3}^2(2- \hat{q_3}+ 2\hat{q_2})  \\ &
+ \hat{q_0}(3p_1 \hat{q_0} - 2 p_1 -\hat{q_0})(p_2 \hat{q_2}^2 + p_3 \hat{q_3}^2),             \end{align}
and
\begin{align}\label{eq:D3}
\nonumber L_2(\mathbf{p}, \hat{\mathbf{q}}) =&~p_2p_3(3p_1^2-2p_1-1)\hat{q_0}^2
 +p_2(5p_1p_2p_3+3p_1p_2-2p_1p_3 +2p_3)\hat{q_2}^2  \\ &\nonumber +p_3(5p_1p_2p_3+3p_1p_3-2p_1p_2+2p_2)\hat{q_3}^2 \\ &           
+2p_2p_3 (1-p_1^2 \hat{q_0} - p_3^2 \hat{q_2} - p_2^2 \hat{q_3}).              
\end{align}
We note that these expressions are exchange invariant, namely, they remain the same under the simultaneous exchange of $p_2$ with $p_3$ and $\hat{q_2}$ with $\hat{q_3}$. 

Next, it is important to implement in Eqs.~\eqref{eq:D1}--\eqref{eq:D3} the relation between $\mathbf{p}$ and $\hat{\mathbf{q}}$ given, for instance, in Eq.~\eqref{eq:C12}, via the Lifshitz-Khalatnikov parameterization of the double-Kasner spacetime described in Appendix C. We recall that for a given $\mathbf{p}$, there are two possible values for $\hat{\mathbf{q}}$, given explicitly by $\hat{\mathbf{q}}'$ and $\hat{\mathbf{q}}''$. The results for the Kretschmann scalar can be expressed as
\begin{equation}\label{eq:D4} 
K(\mathbf{p}, \hat{\mathbf{q}}') = -4p_1p_3\hat{q_0}'\hat{q_2}', \quad K(\mathbf{p}, \hat{\mathbf{q}}'') = -4p_1p_2\hat{q_0}''\hat{q_3}''.
\end{equation} 
In a similar way, one can show that 
\begin{equation}\label{eq:D5} 
L_1(\mathbf{p}, \hat{\mathbf{q}}') = 2p_1 p_3 \hat{q_2}'^2 \hat{q_3}' (3 \hat{q_0}'- \hat{q_3}')        
\end{equation} 
and
\begin{equation}\label{eq:D6} 
L_2(\mathbf{p}, \hat{\mathbf{q}}') = 2p_2 p_3^2 (3 p_1- p_2) \hat{q_0}' \hat{q_2}'.
\end{equation} 
Moreover, we find that
\begin{equation}\label{eq:D7} 
L_1(\mathbf{p}, \hat{\mathbf{q}}'') = 2p_1 p_2  \hat{q_3}''^2 \hat{q_2}'' (3\hat{q_0}''-\hat{q_2}'')        
\end{equation} 
and
\begin{equation}\label{eq:D8} 
L_2(\mathbf{p}, \hat{\mathbf{q}}'') = 2p_3 p_2^2 (3p_1-p_3) \hat{q_0}''\hat{q_3}''.
\end{equation} 
A significant feature of these results is that they are all consistent with the interchange property discussed in Appendix C.

\end{document}